\documentclass[twocolumn, times]{aastex631}

\newcommand{\editR}[2][black]{{\color{#1} #2}}
\newcommand{\editX}[2][black]{{\color{#1}}}

\usepackage{savesym}
\savesymbol{tablenum}
\usepackage{siunitx}
\restoresymbol{SIX}{tablenum}

\usepackage{tipa}
\usepackage{amsmath}
\usepackage{ulem}

\submitjournal{ApJ}

\shorttitle{}
\shortauthors{Arakawa et al.}
\graphicspath{{./}{figures/}}

\begin{document}

\title{Impacts of viscous dissipation on collisional growth and fragmentation of dust aggregates}

\author[0000-0003-0947-9962]{Sota Arakawa}
\affiliation{National Astronomical Observatory of Japan, 2-21-1, Osawa, Mitaka, Tokyo, 181-8588, Japan}
\affiliation{Japan Agency for Marine-Earth Science and Technology, 3173-25, Showa-machi, Kanazawa-ku, Yokohama, 236-0001, Japan}

\author[0000-0001-9659-658X]{Hidekazu Tanaka}
\affiliation{Astronomical Institute, Graduate School of Science Tohoku University, 6-3 Aramaki, Aoba-ku, Sendai 980-8578, Japan}

\author[0000-0002-5486-7828]{Eiichiro Kokubo}
\affiliation{National Astronomical Observatory of Japan, 2-21-1, Osawa, Mitaka, Tokyo, 181-8588, Japan}



\begin{abstract}
Understanding the collisional behavior of dust aggregates consisting of submicron-sized grains is essential to unveiling how planetesimals formed in protoplanetary disks.
It is known that the collisional behavior of individual dust particles strongly depends on the strength of viscous dissipation force; however, impacts of viscous dissipation on the collisional behavior of dust aggregates have not been studied in detail, especially for the cases of oblique collisions.
Here we investigated the impacts of viscous dissipation on the collisional behavior of dust aggregates.
We performed \editX{three-dimensional} numerical simulations of collisions between two equal-mass dust aggregates with various collision velocities and impact parameters.
We also changed the strength of viscous dissipation force systematically.
We found that the threshold collision velocity for the fragmentation of dust aggregates barely depends on the strength of viscous dissipation force when we consider oblique collisions.
In contrast, the size distribution of fragments changes significantly when the viscous dissipation force is considered.
We obtained the empirical fitting formulae for the size distribution of fragments for the case of strong dissipation, which would be useful to study the evolution of size and spatial distributions of dust aggregates in protoplanetary disks.
\end{abstract}



\section{introduction} \label{sec:intro}

Collisional evolution of dust aggregates consisting submicron-sized grains in protoplanetary disks is the first step of the planet formation \citep[e.g.,][]{2014prpl.conf..547J}.
To make planetesimals, dust aggregates should grow through collisions; however, the condition for collisional growth of dust aggregates is still under debate.

In protoplanetary disks, the gas drag transfers the angular momentum and causes the radial drift of dust aggregates.
In the minimum mass solar nebula model \citep{1981PThPS..70...35H}, the dust aggregates drift inwardly with the radial drift velocity of approximately $50~{\rm m}~{\rm s}^{-1}$ at the maximum \citep{1976PThPh..56.1756A, 1977MNRAS.180...57W}.
Since the radial drift velocity depends on the size of dust aggregates, two dust aggregates with different size have a relative velocity due to radial drift motion.
The collision velocity of dust aggregates is also provided by the influence of gas turbulence \citep[e.g.,][]{2007A&A...466..413O, 2021ApJ...911..140S}.
The maximum velocity caused by radial drift and turbulence is typically $10$--$100~{\rm m}~{\rm s}^{-1}$ at $3$--$5~{\rm au}$ from the central star \citep[e.g.,][]{2021ApJ...915...22H}.
Therefore, understanding the collisional behavior of dust aggregates for the collision velocity lower than $100~{\rm m}~{\rm s}^{-1}$ is essential to investigate how dust aggregates evolve into planetesimal in protoplanetary disks.
We note that the strength of gas turbulence would be related to the size and spatial distributions of dust aggregates in protoplanetary disks \citep[e.g.,][]{2012ApJ...753L...8O}, and the size distribution of fragments produced via collisions is important to constrain the collisional growth paths of dust aggregates.

The collisional behavior of dust aggregates has been reported in a large number of studies based on both laboratory experiments \citep[e.g.,][]{2008ARA&A..46...21B, 2010A&A...513A..56G, 2012Icar..221..310S, 2016A&A...593A...3B, 2017A&A...603A..66B, 2018ApJ...853...74S, 2022MNRAS.509.5641S, 2021ApJ...923..134F} and numerical simulations \citep[e.g.,][]{2007ApJ...661..320W, 2008ApJ...677.1296W, 2009ApJ...702.1490W, 2013A&A...559A..62W, 2009A&A...507.1023P, 2013A&A...560A..45S, 2017ApJ...841...36S, 2020A&A...633A..24U, 2021A&A...652A..40U, 2021ApJ...915...22H, 2021ApJ...911..114S}.
It is known that the threshold collision velocity for the fragmentation of dust aggregates, $v_{\rm fra}$, depends on material properties of constituent particles.
For example, numerical simulations by \citet{2009ApJ...702.1490W} and \citet{2021ApJ...915...22H} found that $v_{\rm fra} \simeq 50$--$60~{\rm m}~{\rm s}^{-1}$ when the collision of two equal-mass dust aggregates made of $0.1~\si{\micro}{\rm m}$-sized ${\rm H}_{2} {\rm O}$ ice particles is considered.

The stickiness of dust particles has also been investigated in a huge number of studies \citep[e.g.,][]{2000ApJ...533..454P, 2012arXiv1204.0001G, 2012PThPS.195..101T, 2013JPhD...46Q5303K, 2015ApJ...798...34G, 2017PCCP...1916555N, 2020Icar..35213996N, 2022ApJ...925..173N, 2017ApJ...844..105Q, 2021ApJ...910..130A}.
These studies found that the threshold velocity for the sticking in a head-on collision of individual dust particles is several times higher than that predicted for perfectly elastic spheres.
\citet{2015ApJ...798...34G} and \citet{2021ApJ...910..130A} concluded that the viscous dissipation would play a critical role in collision between micron-sized ${\rm H}_{2} {\rm O}$ ice particles.

Nevertheless, impacts of viscous dissipation on the collisional behavior of dust aggregates have not been studied in detail.
Most of numerical simulations of collisions between dust aggregates did not include the viscous dissipation force \citep[e.g.,][]{1997ApJ...480..647D, 2007ApJ...661..320W, 2008ApJ...677.1296W, 2009ApJ...702.1490W, 2021ApJ...915...22H}.
Although a small number of studies \citep[e.g.,][]{2013A&A...560A..45S, 2021A&A...652A..40U, umstatter2021scaling} introduced the viscous dissipation force to their numerical simulations of collisions between dust aggregates, the dependence of $v_{\rm fra}$ on the strength of viscous dissipation force is still poorly understood.
For head-on collisions between equal-mass dust aggregates, \citet{umstatter2021scaling} reported that the threshold velocity increases with increasing the strength of viscous dissipation force; however, when we consider oblique collisions, whether the viscous dissipation helps the collisional growth of dust aggregates is still unclear.

Here we report the results from numerical simulations of collisions between two equal-mass dust aggregates with various collision velocity and impact parameters.
We also changed the strength of viscous dissipation force systematically.
Surprisingly, we found that the threshold collision velocity for the fragmentation of dust aggregates hardly depends on the strength of viscous dissipation force when we consider oblique collisions.
In contrast, the size distribution of fragments changes significantly when the viscous dissipation force is considered.
We also obtained the empirical fitting formulae for the size distribution of fragments for the case of strong dissipation, which would be useful to study the evolution of size and spatial distributions of dust aggregates in protoplanetary disks.

\section{model}

We performed three-dimensional numerical simulations of collisions between two equal-mass dust aggregates.
The numerical code used in this study is based on the code developed by \citet{2009ApJ...702.1490W}, and we introduced the viscous drag term as \citet{2013A&A...560A..45S} considered.

\subsection{Material parameters}

In this study, we assume that dust particles constituting dust aggregates are made of water ice, and the all particles have the same radius of $r_{1} = 0.1~\si{\micro}{\rm m}$.
The material parameters of water ice particles used in this study are listed in Table \ref{table1}.
These parameters are identical to that used in \citet{2009ApJ...702.1490W} and \citet{2021ApJ...915...22H}.

\begin{table}
\caption{
List of material parameters used in this study.
}
\label{table1}
\centering
\begin{tabular}{ccc}
{\bf Parameter}                  & {\bf Symbol}       & {\bf Value}                   \\ \hline
Particle radius                  & $r_{1}$            & $0.1~\si{\micro}{\rm m}$      \\
Material density                 & $\rho$             & $1000~{\rm kg}~{\rm m}^{-3}$  \\
Surface energy                   & $\gamma$           & $100~{\rm mJ}~{\rm m}^{-2}$   \\
Young's modulus                  & $\mathcal{E}$      & $7~{\rm GPa}$                 \\
Poisson's ratio                   & $\nu$              & $0.25$                        \\
Critical rolling displacement    & $\xi_{\rm crit}$   & $0.8~{\rm nm}$                \\ \hline
\end{tabular}
\end{table}

\subsection{Nomal motion}
\label{sec.normal}

Here we briefly describe the interparticle forces in the normal direction \citep[for details, see][]{1997ApJ...480..647D, 2007ApJ...661..320W}.
When two particles in contact are elastic spheres having a surface energy, their elastic behavior causes a repulsive force and the surface energy causes an attractive force.
\citet{1971RSPSA.324..301J} formulated the particle interaction described below.

Two particles in contact make a contact area.
At the equilibrium state, the contact radius, $a$, is equal to the equilibrium radius, $a_{0}$, given by
\begin{equation}
a_{0} = {\left( \frac{9 \pi \gamma R^{2}}{\mathcal{E}^{*}} \right)}^{1/3},
\end{equation}
where $R = r_{1} / 2$ is the reduced particle radius and $\mathcal{E}^{*} = \mathcal{E} / {[ 2 {( 1 - \nu^2 )} ]}$ is the reduced Young's modulus. 
The contact radius is related to the compression length between two particles in contact, $\delta$, as follows:
\begin{equation}
\frac{\delta}{\delta_{0}} = 3 {\left( \frac{a}{a_{0}} \right)}^{2} - 2 {\left( \frac{a}{a_{0}} \right)}^{1/2},
\end{equation}
where $\delta_{0} = {a_{0}}^{2} / {( 3 R )}$ is the equilibrium compression length when $a = a_{0}$.
Two particles in contact separate when compression length reaches $\delta = - \delta_{\rm c}$, where
\begin{equation}
\label{eq.deltac}
\delta_{\rm c} = {\left( \frac{9}{16} \right)}^{1/3} \delta_{0}.
\end{equation}

In this study, the normal force acting between two particles, $F$, is given by the sum of the two terms;
\begin{equation}
\label{eq.F}
F = F_{\rm E} + F_{\rm D},
\end{equation}
where $F_{\rm E}$ is the force arising from the elastic deformation of particles and $F_{\rm D}$ is the force related to the viscous drag.
The elastic force is given by \citet{1971RSPSA.324..301J} as
\begin{equation}
\frac{F_{\rm E}}{F_{\rm c}} = 4 {\left[ {\left( \frac{a}{a_{0}} \right)}^{3} - {\left( \frac{a}{a_{0}} \right)}^{3/2} \right]},
\end{equation}
where
\begin{equation}
\label{eq.Fc}
F_{\rm c} = 3 \pi \gamma R,
\end{equation}
is the maximum force needed to separate the two particles in contact.
The viscous drag force is given by \citet{2013A&A...560A..45S} as follows;
\begin{equation}
\label{eq.Fd}
F_{\rm D} = \frac{2 T_{\rm vis} \mathcal{E}^{*}}{\nu^{2}} a v_{\rm rel},
\end{equation}
where $v_{\rm rel}$ is the normal component of the relative velocity of the two particles and $T_{\rm vis}$ is the viscoelastic timescale that is related to the strength of the viscous drag force \citep[see also][]{2013JPhD...46Q5303K}.

The value of $T_{\rm vis}$ for $0.1~\si{\micro}{\rm m}$-sized ${\rm H}_{2}{\rm O}$ ice particles is not yet understood.
Based on laboratory experiments \citep{2015ApJ...798...34G, 2016ApJ...827...63M}, the threshold velocity for sticking in a head-on collision of individual micron-sized ice particle is approximately 10 times higher than that predicted from the theoretical model of \citet{1971RSPSA.324..301J}.
To reproduce the experimental results, \citet{2015ApJ...798...34G} showed that $T_{\rm vis} \simeq 10^{-10}~{\rm s}$ is plausible for $r_{1} = 1.5~\si{\micro}{\rm m}$, and \citet{2021ApJ...910..130A} also reported that $T_{\rm vis} \simeq 10^{-8}~{\rm s}$ is plausible for $r_{1} = 90~\si{\micro}{\rm m}$.
A simple extrapolation suggests that $T_{\rm vis} \simeq 10^{-12}$--$10^{-11}~{\rm s}$ for $r_{1} = 0.1~\si{\micro}{\rm m}$.
We note that $T_{\rm vis}$ should depend not only on the particle radius but also on their temperature and the size of crystal domains in reality.
Future studies on the strength of the viscous drag force using molecular dynamics simulations and laboratory experiments are essential.

The potential energy for normal motion of the two particles in contact, $U_{\rm n}$, is given by
\begin{eqnarray}
\frac{U_{\rm n}}{F_{\rm c} \delta_{\rm c}} & = & 4 \times 6^{1/3} \nonumber \\
                                           &   & \times {\left[ \frac{4}{5} {\left( \frac{a}{a_{0}} \right)}^{5} - \frac{4}{3} {\left( \frac{a}{a_{0}} \right)}^{7/2} + \frac{1}{3} {\left( \frac{a}{a_{0}} \right)}^{2} \right]}.
\end{eqnarray}
When two particles collide, the compression length is $\delta = 0$ and the potential energy for normal motion of the newly formed contact is
\begin{equation}
U_{\rm n} {( 0 )} = - \frac{8}{15} \times 2^{2/3} F_{\rm c} \delta_{\rm c},
\end{equation}
and the energy of $- U_{\rm n} {( 0 )}$ dissipates due to connection.
Similarly, when two particles separate, the compression length is $\delta = - \delta_{\rm c}$ and the potential energy for normal motion before the disconnection is
\begin{equation}
U_{\rm n} {( - \delta_{\rm c} )} = \frac{4}{45} F_{\rm c} \delta_{\rm c},
\end{equation}
and the energy of $U_{\rm n} {( - \delta_{\rm c} )}$ dissipates due to disconnection.
In addition, the potential energies stored by tangential displacement also dissipate when two particles separate (see Section \ref{sec.tangential}).
The potential energy at the equilibrium state ($\delta = \delta_{0}$) is
\begin{equation}
U_{\rm n} {( \delta_{0} )} = - \frac{4}{5} \times 6^{1/3} F_{\rm c} \delta_{\rm c}.
\end{equation}
Therefore, the energy needed to break a contact in the equilibrium by quasistatic process ($v_{\rm rel} \to 0$ and $F_{\rm D} \to 0$), $E_{\rm break}$, is given by
\begin{eqnarray}
\label{eq.break}
E_{\rm break} & = & U_{\rm n} {( - \delta_{\rm c} )} - U_{\rm n} {( \delta_{0} )} \nonumber \\
              & = & 6.1 \times 10^{-17}~{\rm J}.
\end{eqnarray}

\subsection{Tangential motion}
\label{sec.tangential}

The tangential motion of two particles in contact is divided into three motions: rolling, sliding, and twisting.
The displacements corresponding to these motions are described as the rotation of two particles in contact.
\citet{2007ApJ...661..320W} described the particle interaction models for these tangential motions \citep[see also][]{1995PMagA..72..783D,1996PMagA..73.1279D}.
Based on the framework of \citet{2007ApJ...661..320W}, tangential motions of particles in contact are described by the linear spring model with critical displacements to their elastic limits.
Here we briefly explain the outline.

\subsubsection{Rolling motion}

The potential energy stored by the rolling displacement is
\begin{equation}
U_{\rm r} = \frac{1}{2} k_{\rm r} \xi^{2},
\end{equation}
where $\xi$ is the length of the rolling displacement and $k_{\rm r}$ is the spring constant that is given by
\begin{equation}
k_{\rm r} = \frac{4 F_{\rm c}}{R}.
\end{equation}
The critical rolling displacement is set to be $\xi_{\rm crit} = 0.8~{\rm nm}$.\footnote{
We note that $\xi_{\rm crit}$ represents the adhesion hysteresis of the contact area, and $\xi_{\rm crit}$ may depend on the crack velocity \citep[see][and references therein]{2014JPhD...47q5302K}.
Although we do not consider the viscoelastic model for the adhesion hysteresis of the contact area \citep[e.g.,][]{2021ApJ...910..130A} for simplicity, future studies on this point would be interesting.
}
The critical energy required to start rolling, $E_{\rm r, crit}$, is given by 
\begin{eqnarray}
\label{eq.crit-roll}
E_{\rm r, crit} & = & \frac{1}{2} k_{\rm r} {\xi_{\rm crit}}^{2} \nonumber \\
                & = & 1.2 \times 10^{-18}~{\rm J}.
\end{eqnarray}
\citet{2007ApJ...661..320W} introduced $E_{\rm roll}$ as the energy needed to rotate a particle by $\pi / 2$ radian around its contact point \citep[see also][]{1997ApJ...480..647D}.
For the case of water ice particles of $r_{1} = 0.1~\si{\micro}{\rm m}$, $E_{\rm roll}$ is given by
\begin{eqnarray}
\label{eq.roll}
E_{\rm roll} & = & k_{\rm r} \xi_{\rm crit} \pi R \nonumber \\
             & = & 4.7 \times 10^{-16}~{\rm J}.
\end{eqnarray}

\subsubsection{Sliding motion}

The potential energy stored by the sliding displacement is
\begin{equation}
U_{\rm s} = \frac{1}{2} k_{\rm s} \zeta^{2},
\end{equation}
where $\zeta$ is the length of the sliding displacement and $k_{\rm s}$ is the spring constant that is given by
\begin{equation}
k_{\rm s} = 8 a_{0} \mathcal{G}^{\star},
\end{equation}
where $\mathcal{G}^{\star} = \mathcal{G} / {[ 2 {( 2 - \nu )} ]}$.
The shear modulus, $\mathcal{G}$, is related to the Young's modulus and the Poisson's ratio as $\mathcal{G} = \mathcal{E} / {[ 2 {( 1 + \nu )} ]}$.
The critical sliding displacement, $\zeta_{\rm crit}$, is 
\begin{equation}
\zeta_{\rm crit} = \frac{2 - \nu}{16 \pi} a_{0}.
\end{equation}
The critical energy required to start sliding, $E_{\rm s, crit}$, is given by 
\begin{eqnarray}
\label{eq.crit-slide}
E_{\rm s, crit} & = & \frac{1}{2} k_{\rm s} {\zeta_{\rm crit}}^{2} \nonumber \\
                & = & 7.3 \times 10^{-18}~{\rm J},
\end{eqnarray}
and the energy needed to slide a particle by $\pi / 2$ radian around its contact point, $E_{\rm slide}$, is given by
\begin{eqnarray}
\label{eq.slide}
E_{\rm slide} & = & k_{\rm s} \zeta_{\rm crit} \pi R \nonumber \\
              & = & 1.7 \times 10^{-14}~{\rm J}.
\end{eqnarray}

\subsubsection{Twisting motion}

The potential energy stored at a twisting angle $\phi$ is
\begin{equation}
U_{\rm t} = \frac{1}{2} k_{\rm r} \phi^{2},
\end{equation}
and the spring constant $k_{\rm t}$ is given by
\begin{equation}
k_{\rm t} = \frac{16}{3} \mathcal{G}^{'} {a_{0}}^{3},
\end{equation}
where $\mathcal{G}^{'} = \mathcal{G} / 2$ is the reduced shear modulus.
The critical angle for twisting, $\phi_{\rm crit}$, is 
\begin{equation}
\phi_{\rm crit} = \frac{1}{16 \pi}.
\end{equation}
The critical energy required to start twisting, $E_{\rm t, crit}$, is given by 
\begin{eqnarray}
\label{eq.crit-twist}
E_{\rm t, crit} & = & \frac{1}{2} k_{\rm t} {\phi_{\rm crit}}^{2} \nonumber \\
                & = & 2.8 \times 10^{-18}~{\rm J},
\end{eqnarray}
and the energy needed to twist over $\pi / 2$ radian is given by 
\begin{eqnarray}
\label{eq.twist}
E_{\rm twist} & = & k_{\rm t} \phi_{\rm crit} \frac{\pi}{2} \nonumber \\
              & = & 4.4 \times 10^{-16}~{\rm J}.
\end{eqnarray}

In our numerical simulations, the energy is normalized by $F_{\rm c} \delta_{\rm c} = 4.0 \times 10^{-17}~{\rm J}$.
The particle interaction energies for both normal and tangential motions are listed in Table \ref{table2}.

\begin{table}
\caption{
List of particle interaction energies.
}
\label{table2}
\centering
\begin{tabular}{ccc}
{\bf Symbol}                  & {\bf Value}                    & {\bf Equation}   \\ \hline
$F_{\rm c} \delta_{\rm c}$    & $4.0 \times 10^{-17}~{\rm J}$  & Eqs.~{(\ref{eq.deltac})} and {(\ref{eq.Fc})}  \\
$E_{\rm break}$               & $6.1 \times 10^{-17}~{\rm J}$  & Eq.~{(\ref{eq.break})}  \\
$E_{\rm r, crit}$             & $1.2 \times 10^{-18}~{\rm J}$  & Eq.~{(\ref{eq.crit-roll})}   \\
$E_{\rm roll}$                & $4.7 \times 10^{-16}~{\rm J}$  & Eq.~{(\ref{eq.roll})}   \\
$E_{\rm s, crit}$             & $7.3 \times 10^{-18}~{\rm J}$  & Eq.~{(\ref{eq.crit-slide})}  \\
$E_{\rm slide}$               & $1.7 \times 10^{-14}~{\rm J}$  & Eq.~{(\ref{eq.slide})}  \\
$E_{\rm t, crit}$             & $2.8 \times 10^{-18}~{\rm J}$  & Eq.~{(\ref{eq.crit-twist})}  \\
$E_{\rm twist}$               & $4.4 \times 10^{-16}~{\rm J}$  & Eq.~{(\ref{eq.twist})}  \\ \hline
\end{tabular}
\end{table}

\subsection{Initial condition}

As the initial condition, we prepare two dust aggregates before collisions by ballistic particle--cluster aggregation (BPCA), i.e., sequential hit-and-stick collisions between a cluster (aggregate) and multiple single particles \citep[e.g.,][]{1992A&A...262..315M}.
In this study, we set the number of particles constituting the target aggregate, $N_{\rm tar}$, is equal to that for the projectile aggregate, $N_{\rm pro}$, and both $N_{\rm tar}$ and $N_{\rm pro}$ are set to be 50,000.
The total number of particles in a simulation, $N_{\rm tot} = N_{\rm tar} + N_{\rm pro}$, is therefore 100,000.

Offset collisions of two dust aggregates is described by the impact parameter, $b_{\rm off}$.
The maximum impact parameter, $b_{\rm max}$, is the sum of the radii of target and projectile aggregates; $b_{\rm max} = r_{\rm c, tar} + r_{\rm c, pro}$.
The radii of dust aggregates are calculate by the characteristic radius, $r_{\rm c}$, which is given by $r_{\rm c} = \sqrt{5/3} r_{\rm gyr}$, and $r_{\rm gyr}$ is the gyration radius \citep{1992A&A...262..315M}.
The radii of two dust aggregates used in this study are $r_{\rm c, tar} = 72.29 r_{1}$ and $r_{\rm c, pro} = 72.25 r_{1}$.
Hereafter we use the normalized impact parameter, $B_{\rm off}$, which is defined as
\begin{equation}
B_{\rm off} \equiv \frac{b_{\rm off}}{b_{\rm max}}.
\end{equation}

The square of the normalized impact parameter ranges from ${B_{\rm off}}^{2} = 0$ to $1$ with an interval of $1/12$.
The collision velocity of two dust aggregates, $v_{\rm col}$, is set to $10^{( 0.1 i )}~{\rm m}~{\rm s}^{-1}$, where $i = 13$, $14$, \ldots, $20$.
We also change $T_{\rm vis}$ that controls the strength of the viscous dissipation.
We set $T_{\rm vis} = 0~{\rm ps}$, $1~{\rm ps}$, $3~{\rm ps}$, and $6~{\rm ps}$ ($1~{\rm ps} = 10^{-12}~{\rm s}$).
Therefore we carried out 416 ($= 13 \times 8 \times 4$) runs in this study.

\section{Results}

\subsection{Examples of collisional outcomes}

\begin{figure*}
\centering
\includegraphics[width=\textwidth]{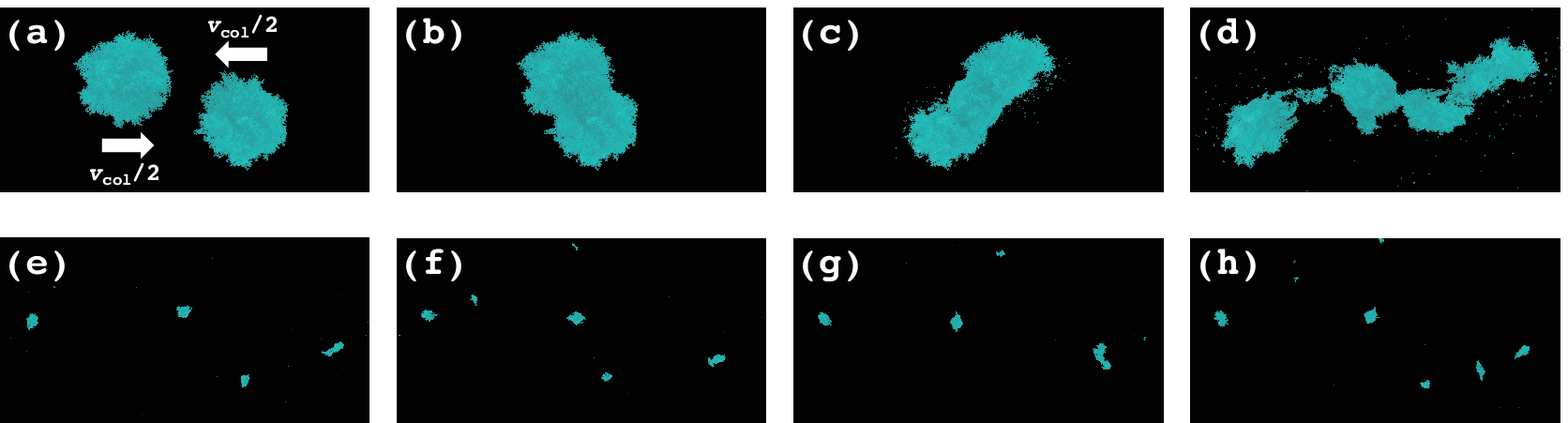}
\caption{
Snapshots of collisional outcomes \editX{with different $T_{\rm vis}$}.
Here we set $v_{\rm col} = 50.1~{\rm m}~{\rm s}^{-1}$ and ${B_{\rm off}}^{2} = 3/12$.
\editR{Panels (a)--(d) are snapshots for $T_{\rm vis} = 0~{\rm ps}$ (at $t = 0~\si{\micro}{\rm s}$, $0.32~\si{\micro}{\rm s}$, $0.63~\si{\micro}{\rm s}$, and $1.27~\si{\micro}{\rm s}$, respectively).
Panels (e)--(h) are snapshots at $t = 6.34~\si{\micro}{\rm s}$ (for $T_{\rm vis} = 0~{\rm ps}$, $1~{\rm ps}$, $3~{\rm ps}$, and $6~{\rm ps}$, respectively).
}}
\label{fig.1}
\end{figure*}

Figure \ref{fig.1} shows the snapshots of collisional outcomes \editX{with different $T_{\rm vis}$}.
Here we show the results for the cases of $v_{\rm col} = 50.1~{\rm m}~{\rm s}^{-1}$ and ${B_{\rm off}}^{2} = 3/12$.
\editX{Panels (a)--(e) show the results for the case of $T_{\rm vis} = 0~{\rm ps}$, panels (f)--(j) show the results for the case of $T_{\rm vis} = 1~{\rm ps}$, panels (k)--(o) show the results for the case of $T_{\rm vis} = 3~{\rm ps}$, and panels (p)--(t) show the results for the case of $T_{\rm vis} = 6~{\rm ps}$.}
\editX{As shown in Figure \ref{fig.1},} \editR{We found that} the collisional behavior is similar among different settings of the viscous dissipation when we focus on large fragments \editR{(panels (e)--(h))}.
In addition, the number of constituent particles in the largest fragment, $N_{\rm lar}$, hardly depends on $T_{\rm vis}$: $N_{\rm lar} = 43957$ for $T_{\rm vis} = 0~{\rm ps}$, $N_{\rm lar} = 44927$ for $T_{\rm vis} = 1~{\rm ps}$, $N_{\rm lar} = 46263$ for $T_{\rm vis} = 3~{\rm ps}$, and $N_{\rm lar} = 47919$ for $T_{\rm vis} = 6~{\rm ps}$.
The amount of small fragments, however, strongly depends on $T_{\rm vis}$.
A large number of small fragments composed of one or two particle(s) are shown in Figures \ref{fig.1}(c) and \ref{fig.1}(d).
In contrast, the number of small fragments significantly decreases when we non-zero $T_{\rm vis}$ is considered.
In the following part of this section, we discuss the effects of the viscous dissipation on the collisional behavior of dust aggregates quantitatively.

\subsection{Growth efficiency}

First, we discuss the parameter dependence of the largest fragment formed after a collision.
Here we introduce the collisional growth efficiency, $f_{\rm gro}$, which is defined by \citet{2013A&A...559A..62W} as follows:
\begin{equation}
f_{\rm gro} \equiv \frac{N_{\rm lar} - N_{\rm tar}}{N_{\rm pro}}.
\end{equation}

\begin{figure*}
\centering
\includegraphics[width=0.48\textwidth]{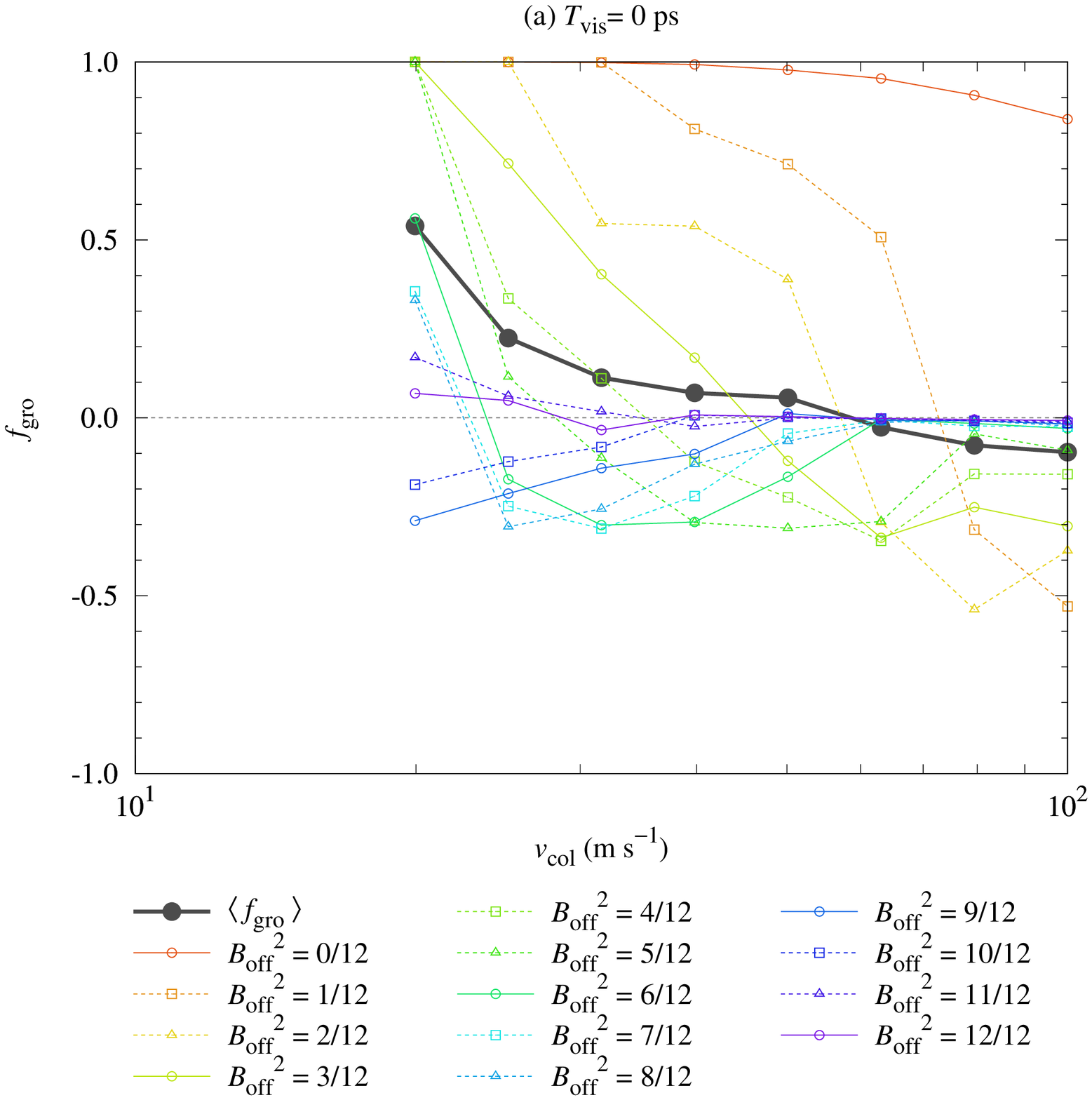}
\includegraphics[width=0.48\textwidth]{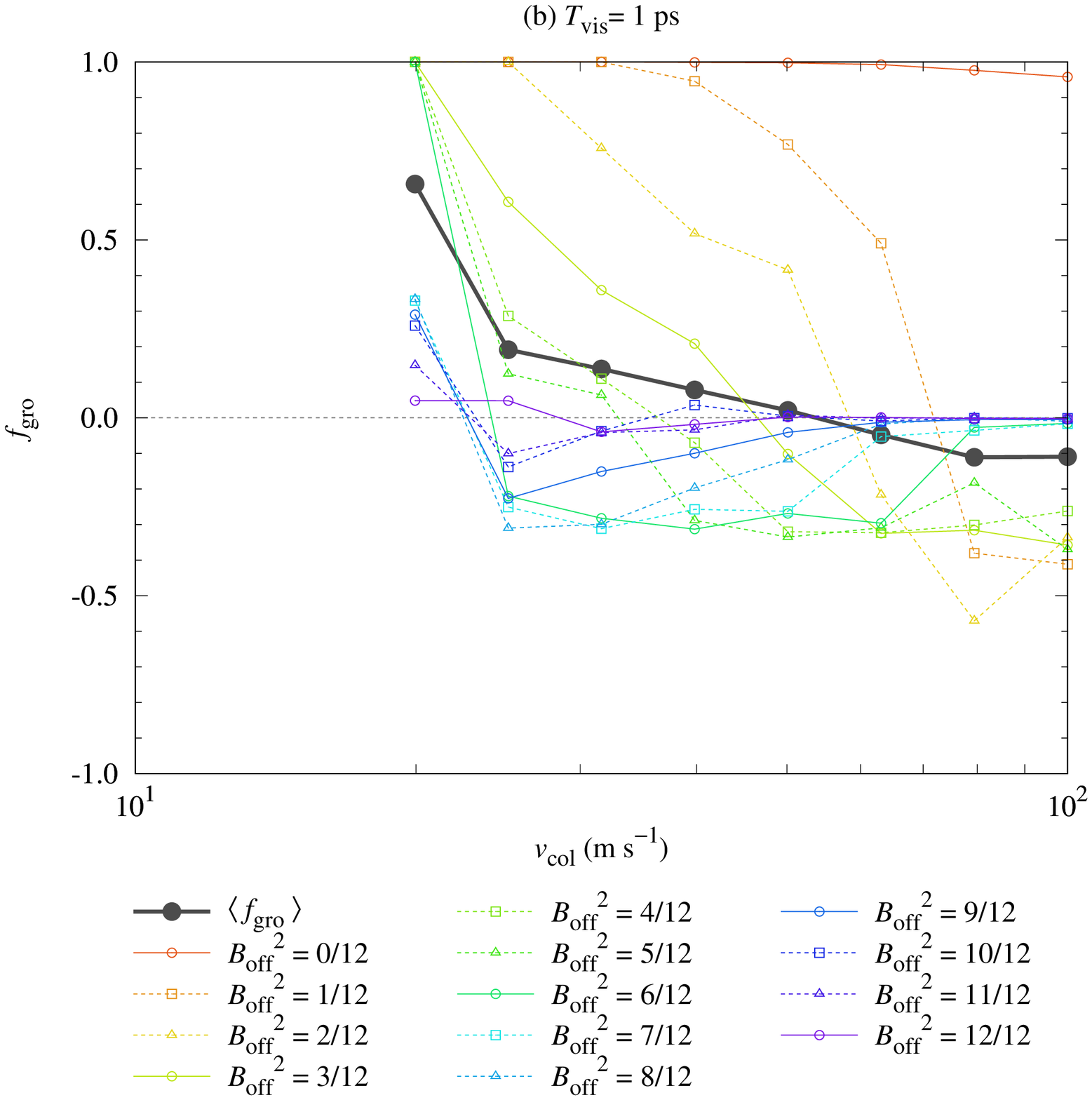}
\includegraphics[width=0.48\textwidth]{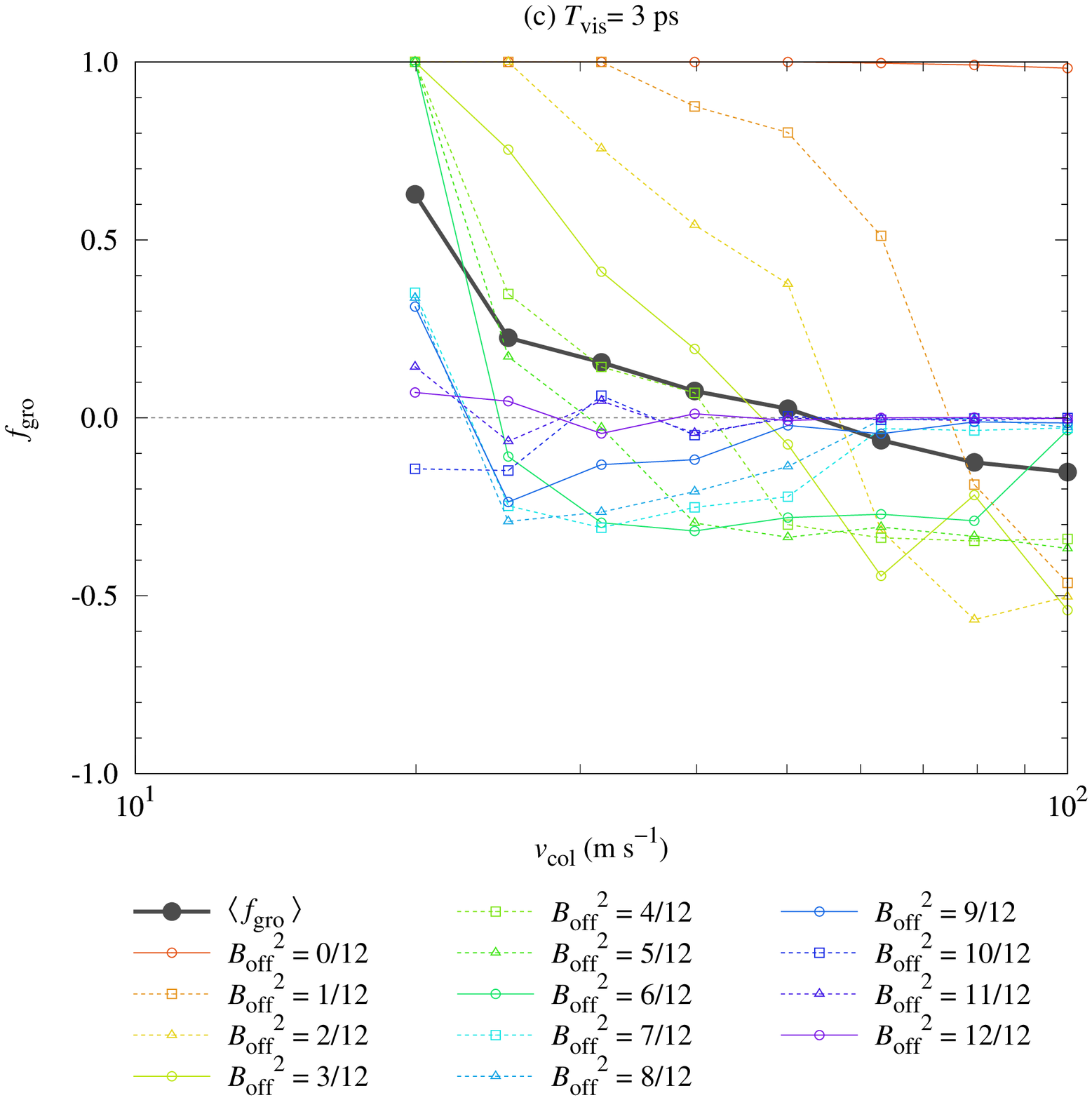}
\includegraphics[width=0.48\textwidth]{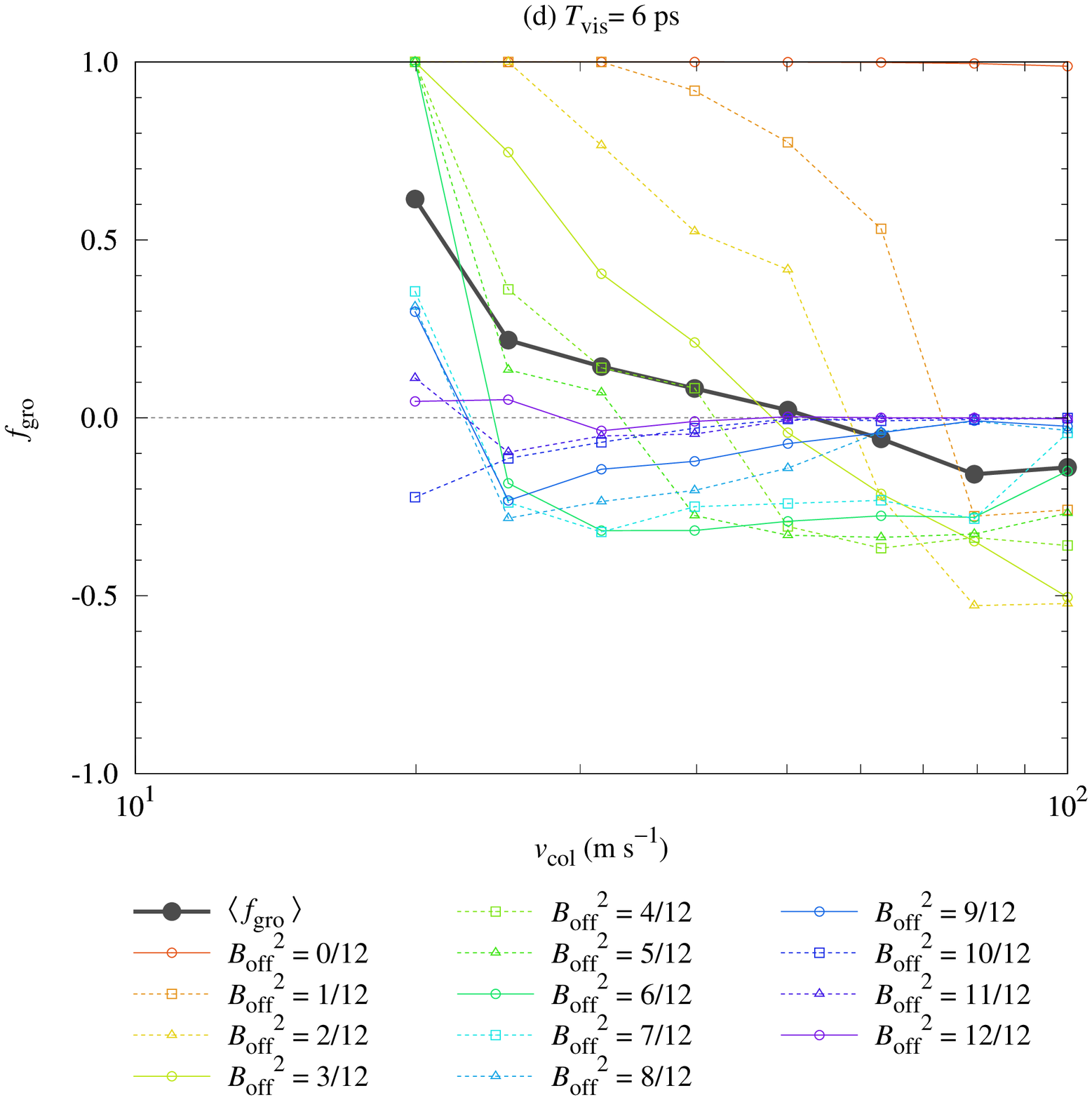}
\caption{
Collisional growth efficiency, $f_{\rm gro}$, for different settings of $T_{\rm vis}$, $B_{\rm off}$, and $v_{\rm col}$.
}
\label{fig.fgro}
\end{figure*}

Figure \ref{fig.fgro} shows the collisional growth efficiency.
In the range of $20~{\rm m}~{\rm s}^{-1} < v_{\rm col} < 100~{\rm m}~{\rm s}^{-1}$, $f_{\rm gro} \simeq 1$ for the cases of head-on collisions (i.e., $B_{\rm off} = 0$).
In contrast, we found that $f_{\rm gro}$ is negative for several cases with non-zero $B_{\rm off}$.
We can observe these trends in both zero and non-zero $T_{\rm vis}$ cases as shown in Figure \ref{fig.fgro}.

When we consider the collisions between dust aggregates in protoplanetary disks, the impact parameter varies with each collision event. 
Then the average value of $f_{\rm gro}$ weighted over $B_{\rm off}$ would be the measure to investigating the dust growth in protoplanetary disks.
The average value of a variable $A$ weighted over $B_{\rm off}$ is given by
\begin{equation}
{\langle A \rangle} \equiv \frac{1}{\pi} \int_{0}^{1} {\rm d}{B_{\rm off}}~2 \pi B_{\rm off} {A (B_{\rm off})}.
\end{equation}
The gray lines in Figure \ref{fig.fgro} shows the $B_{\rm off}$-weighted average of $f_{\rm gro}$, i.e., ${\langle f_{\rm gro} \rangle}$.
As ${\langle f_{\rm gro} \rangle}$ decreases with increasing $v_{\rm col}$, we can calculate the threshold collision velocity for the fragmentation of dust aggregates, $v_{\rm fra}$, which is defined as the velocity where ${\langle f_{\rm gro} \rangle} = 0$.

\begin{figure}
\centering
\includegraphics[width=\columnwidth]{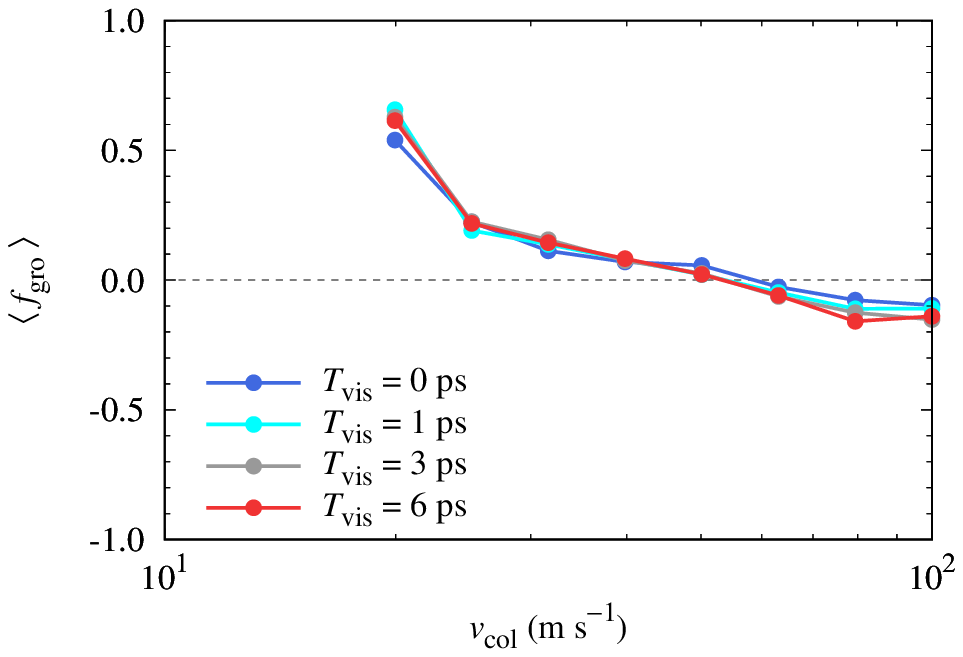}
\caption{
$B_{\rm off}$-weighted collisional growth efficiency, ${\langle f_{\rm gro} \rangle}$.
}
\label{fig.fgro-ave}
\end{figure}

Figure \ref{fig.fgro-ave} shows ${\langle f_{\rm gro} \rangle}$ as a function of $v_{\rm col}$.
We found that ${\langle f_{\rm gro} \rangle}$ hardly depends on the strength of the viscous dissipation, $T_{\rm vis}$.
For all cases of $T_{\rm vis}$, ${\langle f_{\rm gro} \rangle}$ is positive for $v_{\rm col} < 50~{\rm m}~{\rm s}^{-1}$ and is negative for $v_{\rm col} > 60~{\rm m}~{\rm s}^{-1}$.
Therefore $v_{\rm fra} \simeq 50$--$60~{\rm m}~{\rm s}^{-1}$ is obtained from our numerical simulations; this value is in good agreement with that obtained from previous studies \citep{2009ApJ...702.1490W, 2021ApJ...915...22H}.

\begin{figure}
\centering
\includegraphics[width=\columnwidth]{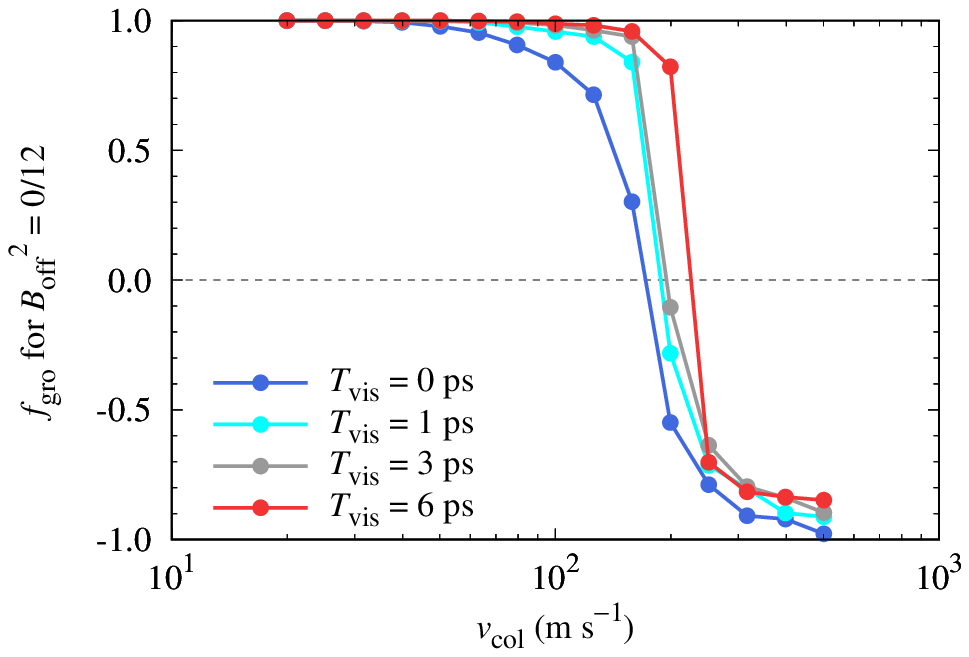}
\caption{
Collisional growth efficiency for high-speed head-on collisions.
}
\label{fig.fgro-headon}
\end{figure}

We note that $f_{\rm gro}$ for head-on collisions strongly depends on $T_{\rm vis}$ as \citet{umstatter2021scaling} reported.
Figure \ref{fig.fgro-headon} shows $f_{\rm gro}$ for high-speed head-on collisions.
We confirmed that the threshold velocity for collisional growth/fragmentation for head-on collisions increases with $T_{\rm vis}$, although ${\langle f_{\rm gro} \rangle}$ should be considered for the dust growth in protoplanetary disks.

\subsection{Threshold velocity for growth/fragmentation}
\label{sec.vfrag}

The threshold velocity for sticking in a head-on collision of individual dust particle, $v_{\rm stick}$, is a function of $T_{\rm vis}$.\footnote{
When we consider a head-on collision of two individual dust particles without fragmentation/evaporation/melting, the collisional outcomes are classified into two cases: sticking (i.e., $f_{\rm gro} = 1$) and bouncing (i.e., $f_{\rm gro} = 0$).
The collisional outcome is sticking when $v_{\rm col} \le v_{\rm stick}$, and bouncing occurs if $v_{\rm col} > v_{\rm stick}$.
}
For a head-on collision of identical particles, the equation of motion is given by
\begin{equation}
\label{eq.dddelta}
\ddot{\delta} = - \frac{F}{m^{*}},
\end{equation}
where $m^{*} = m_{1} / 2$ is the reduced mass, and $m_{1} = {( 4 \pi / 3 )} \rho {r_{1}}^{3}$ is the mass of each particle.
The normal force acting between two particles, $F$, is given by Eq.~{(\ref{eq.F})}, and we integrated Eq.~{(\ref{eq.dddelta})} for different settings of $T_{\rm vis}$ and $v_{\rm col}$.
Then we obtained the threshold velocity for sticking as a function of $T_{\rm vis}$.

\begin{figure}
\centering
\includegraphics[width=\columnwidth]{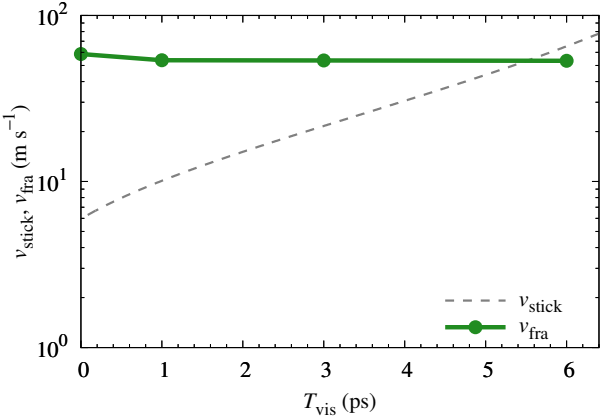}
\caption{
$v_{\rm stick}$ and $v_{\rm fra}$ as functions of $T_{\rm vis}$.
}
\label{fig.Vst}
\end{figure}

Figure \ref{fig.Vst} shows $v_{\rm stick}$ and $v_{\rm fra}$ as functions of $T_{\rm vis}$.
Although $v_{\rm stick}$ for $T_{\rm vis} = 6~{\rm ps}$ is approximately 10 times higher than that for $T_{\rm vis} = 0~{\rm ps}$, $v_{\rm fra}$ hardly depends on  $T_{\rm vis}$.
This indicates that the dominant energy dissipation process for collisions of dust aggregates is different from that for head-on collisions of individual particles.

\subsection{Size distribution of fragments}

\begin{figure*}
\centering
\includegraphics[width=0.48\textwidth]{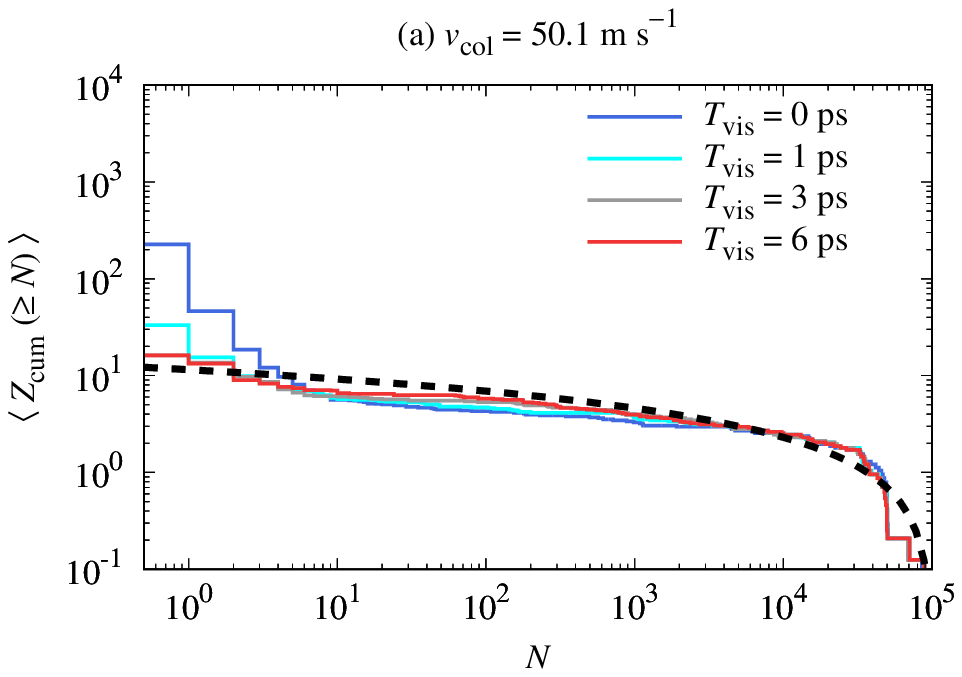}
\includegraphics[width=0.48\textwidth]{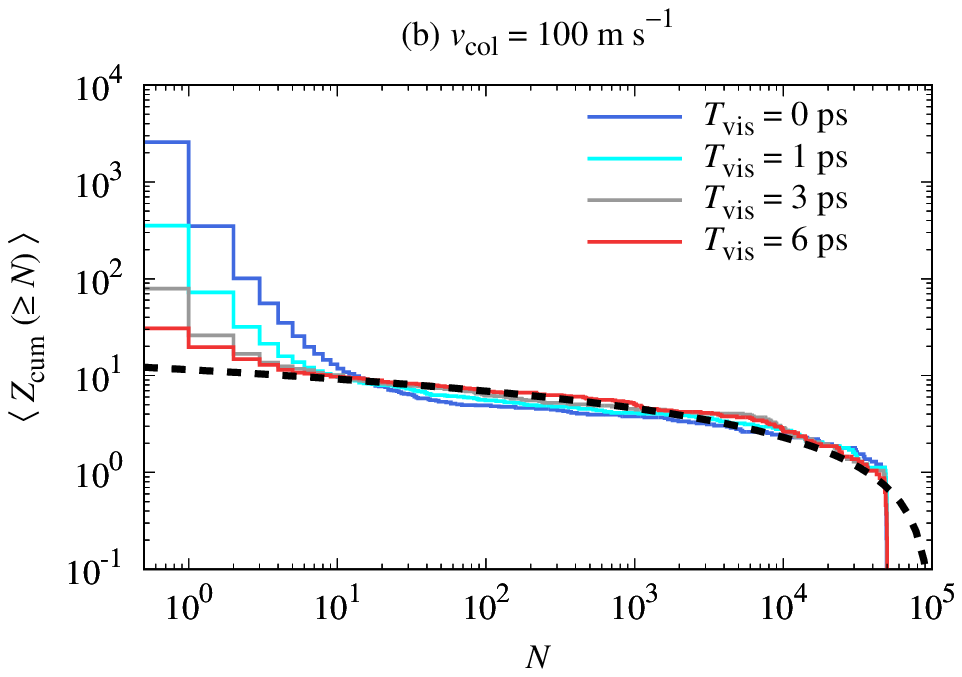}
\caption{
$B_{\rm off}$-weighted average of ${Z_{\rm cum} (\ge N)}$, ${\langle Z_{\rm cum} (\ge N) \rangle}$.
(a) For the case of $v_{\rm col} = 50.1~{\rm m}~{\rm s}^{-1}$.
(b) For the case of $v_{\rm col} = 100~{\rm m}~{\rm s}^{-1}$.
Black dashed lines are the empirical fittings (see Section \ref{sec.emp}).
}
\label{fig.Zcum}
\end{figure*}

Second, we report the size distribution of fragments formed after collisions.
We define ${n (N)}$ as the number of fragments that consist of $N$ particles.
The cumulative number of fragments that contain not smaller than $N$ particles, ${Z_{\rm cum} (\ge N)}$, is defined as follows:
\begin{equation}
{Z_{\rm cum} (\ge N)} \equiv \sum_{N' = {\lceil N \rceil}}^{N_{\rm tot}}\ {n (N')},
\end{equation}
where ${\lceil N \rceil}$ is the smallest integer that is not smaller than $N$.
Figure \ref{fig.Zcum} shows the $B_{\rm off}$-weighted average of ${Z_{\rm cum} (\ge N)}$ for the cases of $v_{\rm col} = 50.1~{\rm m}~{\rm s}^{-1}$ and $100~{\rm m}~{\rm s}^{-1}$.
We found that ${\langle Z_{\rm cum} (\ge N) \rangle}$ strongly depends on $T_{\rm vis}$ for $N \lesssim 10$.
In contrast, the dependence of ${\langle Z_{\rm cum} (\ge N) \rangle}$ on $T_{\rm vis}$ is weak for $N \gg 10$.
This is consistent with the fact that the collisional behavior is similar among different settings of $T_{\rm vis}$ when we focus on large fragments (see Figure \ref{fig.1}).
We also note that ${\langle Z_{\rm cum} (\ge N) \rangle}$ depends on $v_{\rm col}$, especially for the cases of $T_{\rm vis} = 0~{\rm ps}$.
This trend is consistent with the results of \citet{2009ApJ...702.1490W}.

\begin{figure*}
\centering
\includegraphics[width=0.48\textwidth]{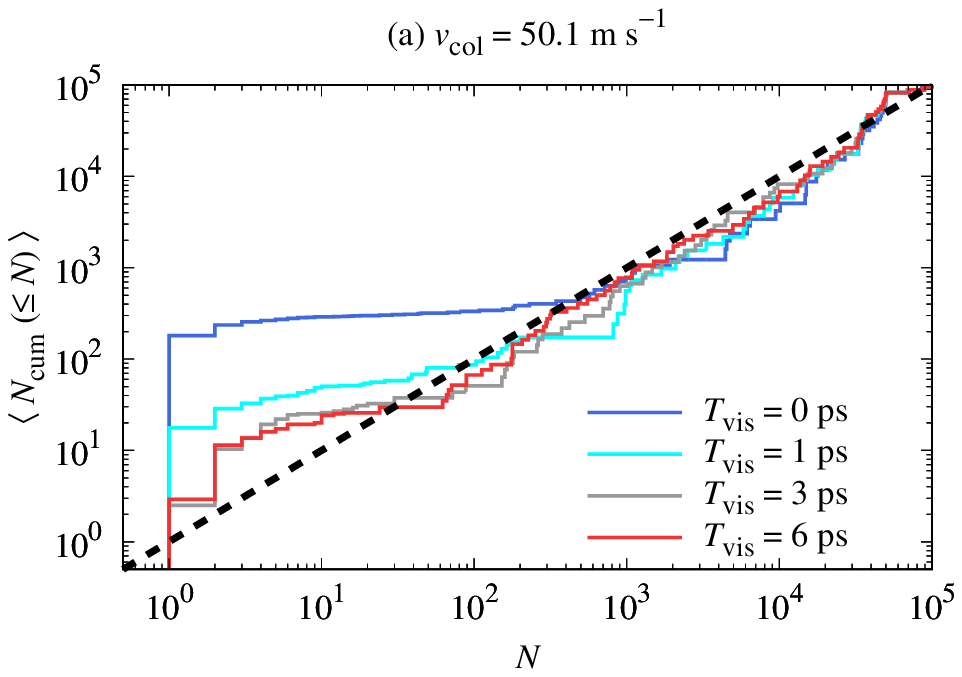}
\includegraphics[width=0.48\textwidth]{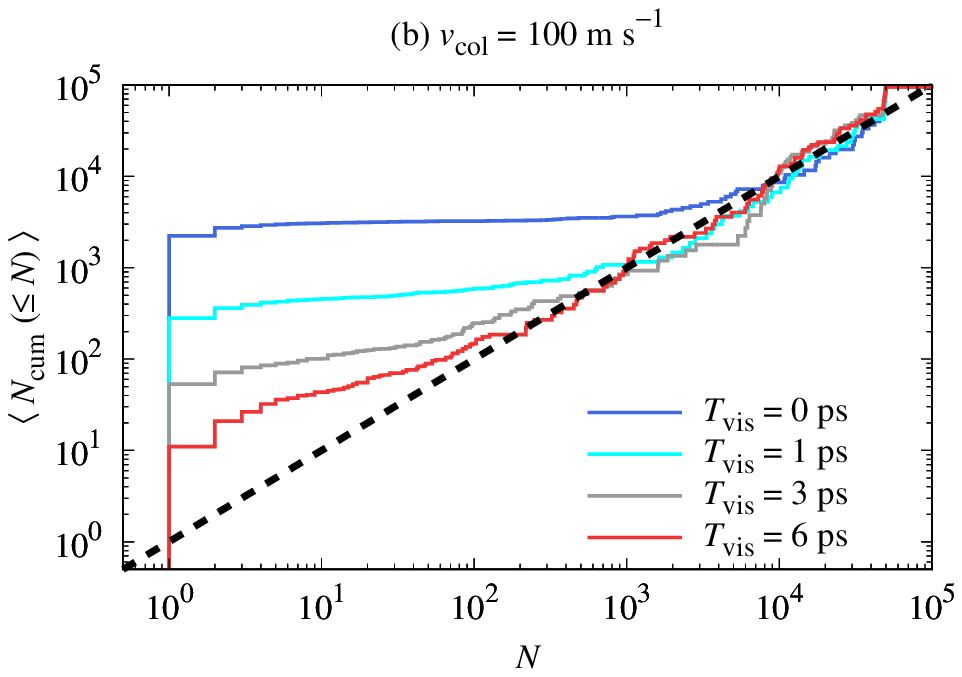}
\caption{
$B_{\rm off}$-weighted average of ${N_{\rm cum} (\le N)}$, ${\langle N_{\rm cum} (\le N) \rangle}$.
(a) For the case of $v_{\rm col} = 50.1~{\rm m}~{\rm s}^{-1}$.
(b) For the case of $v_{\rm col} = 100~{\rm m}~{\rm s}^{-1}$.
Black dashed lines are the empirical fittings (see Section \ref{sec.emp}).
}
\label{fig.Ncum}
\end{figure*}

The cumulative number of particles that are constituents of fragments that contain not larger than $N$ particles, ${N_{\rm cum} (\le N)}$, is defined as follows:
\begin{equation}
{N_{\rm cum} (\le N)} \equiv \sum_{N' = 1}^{{\lfloor N \rfloor}}\ {n (N')} N',
\end{equation}
where ${\lfloor N \rfloor}$ is the largest integer that is not larger than $N$.
Figure \ref{fig.Ncum} shows the $B_{\rm off}$-weighted average of ${N_{\rm cum} (\le N)}$ for the cases of $v_{\rm col} = 50.1~{\rm m}~{\rm s}^{-1}$ and $100~{\rm m}~{\rm s}^{-1}$.
As shown in Figure \ref{fig.Ncum}, largest fragments dominate the mass for the cases of both $v_{\rm col} = 50.1~{\rm m}~{\rm s}^{-1}$ and $100~{\rm m}~{\rm s}^{-1}$.
In addition, our numerical results indicate that ${\langle N_{\rm cum} (\le N) \rangle}$ is approximately proportional to $N$ for $N \gg 10^{2}$ when we consider the non-zero $T_{\rm vis}$.
We provide an empirical fitting of ${\langle N_{\rm cum} (\le N) \rangle}$ in Section \ref{sec.emp}.

\begin{figure*}
\centering
\includegraphics[width=0.48\textwidth]{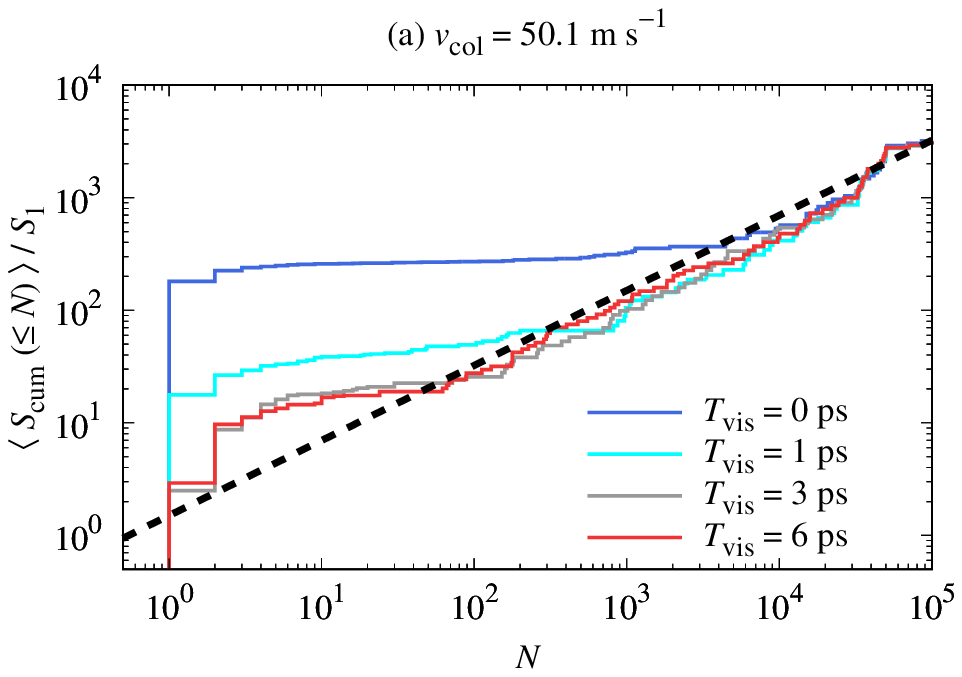}
\includegraphics[width=0.48\textwidth]{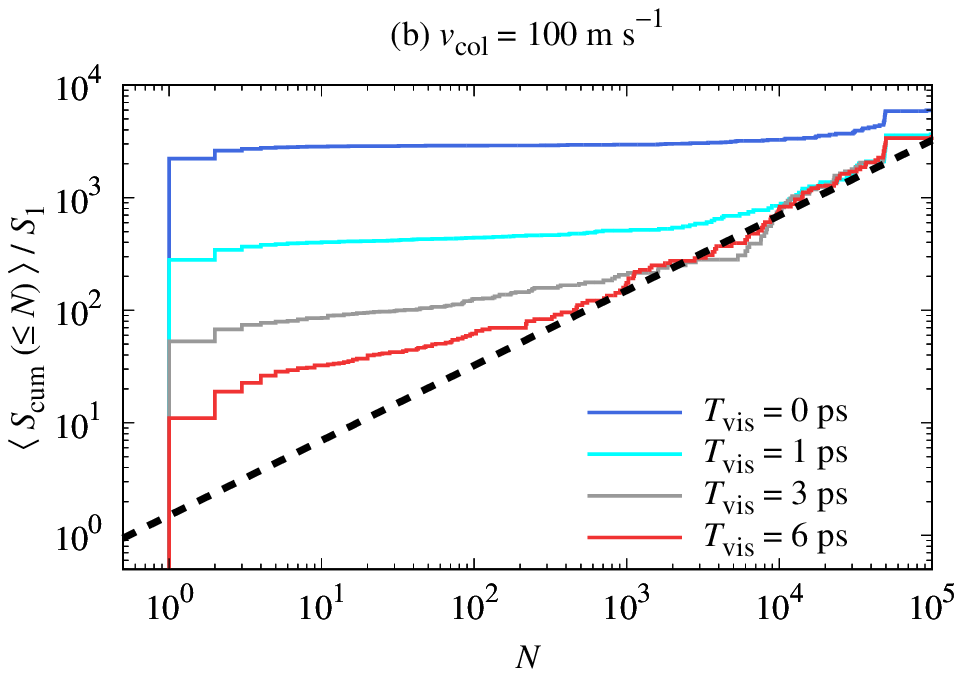}
\caption{
$B_{\rm off}$-weighted average of ${S_{\rm cum} (\le N)}$, ${\langle S_{\rm cum} (\le N) \rangle}$.
(a) For the case of $v_{\rm col} = 50.1~{\rm m}~{\rm s}^{-1}$.
(b) For the case of $v_{\rm col} = 100~{\rm m}~{\rm s}^{-1}$.
Black dashed lines are the empirical fittings (see Section \ref{sec.emp}).
}
\label{fig.Scum}
\end{figure*}

We evaluate the contribution to the geometric cross section.
The geometric cross section of single particle, $S_{1}$, is given by $S_{1} = \pi {r_{1}}^{2}$.
We define the equivalent geometric cross section of a fragment that consist of $N$ particles, ${S (N)}$, as follows: ${S (N)} \equiv \pi {r_{1}}^{2} N^{2/3}$ .
Then the cumulative of the geometric cross section of a fragment that contain not larger than $N$ particles, ${S_{\rm cum} (\le N)}$, is given by
\begin{equation}
{S_{\rm cum} (\le N)} \equiv \sum_{N' = 1}^{{\lfloor N \rfloor}}\ {n (N')} {S (N')}.
\end{equation}
Figure \ref{fig.Scum} shows the $B_{\rm off}$-weighted average of ${S_{\rm cum} (\le N)}$ for the cases of $v_{\rm col} = 50.1~{\rm m}~{\rm s}^{-1}$ and $100~{\rm m}~{\rm s}^{-1}$.
We found that ${\langle S_{\rm cum} (\le N_{\rm tot}) \rangle}$ is dominated by largest fragments except for the case of $v_{\rm col} = 100~{\rm m}~{\rm s}^{-1}$ and $T_{\rm vis} = 0~{\rm ps}$ (blue line of Figure \ref{fig.Scum}(b)).

\section{Empirical fittings for size distribution of fragments}
\label{sec.emp}

\begin{figure}
\centering
\includegraphics[width=0.48\textwidth]{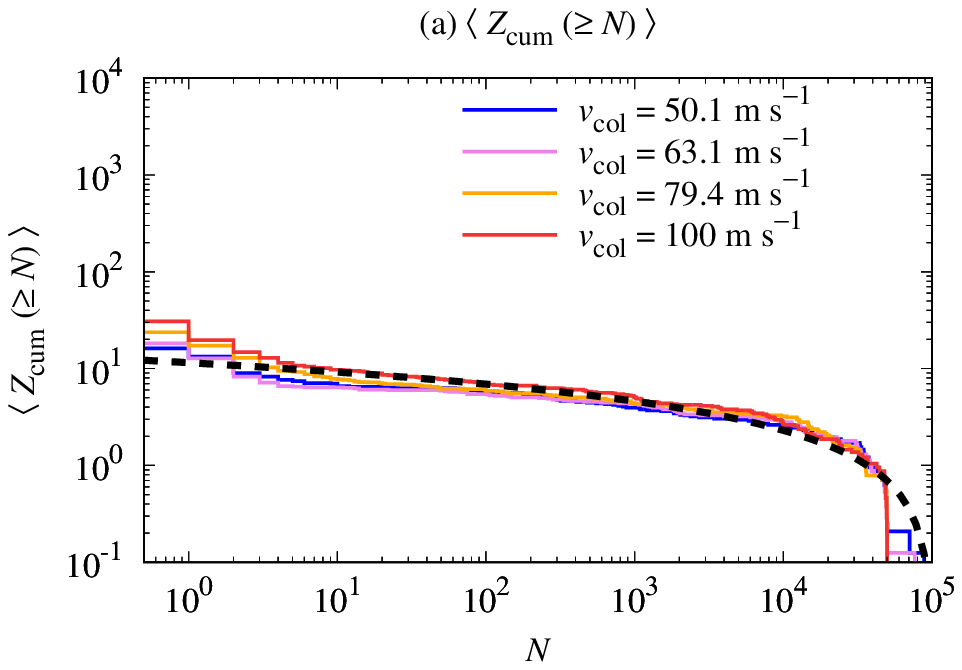}
\includegraphics[width=0.48\textwidth]{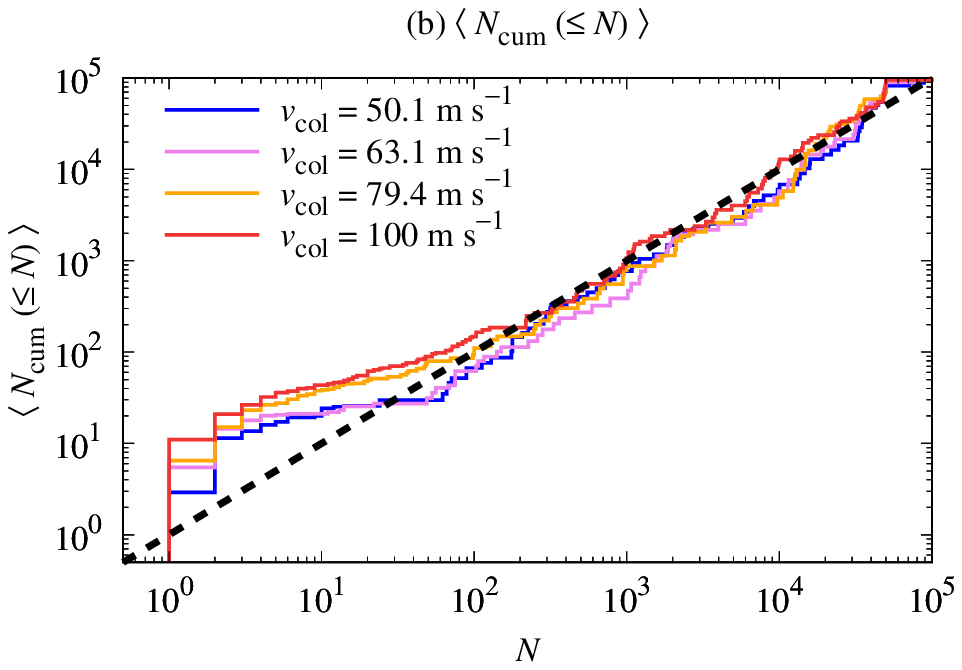}
\includegraphics[width=0.48\textwidth]{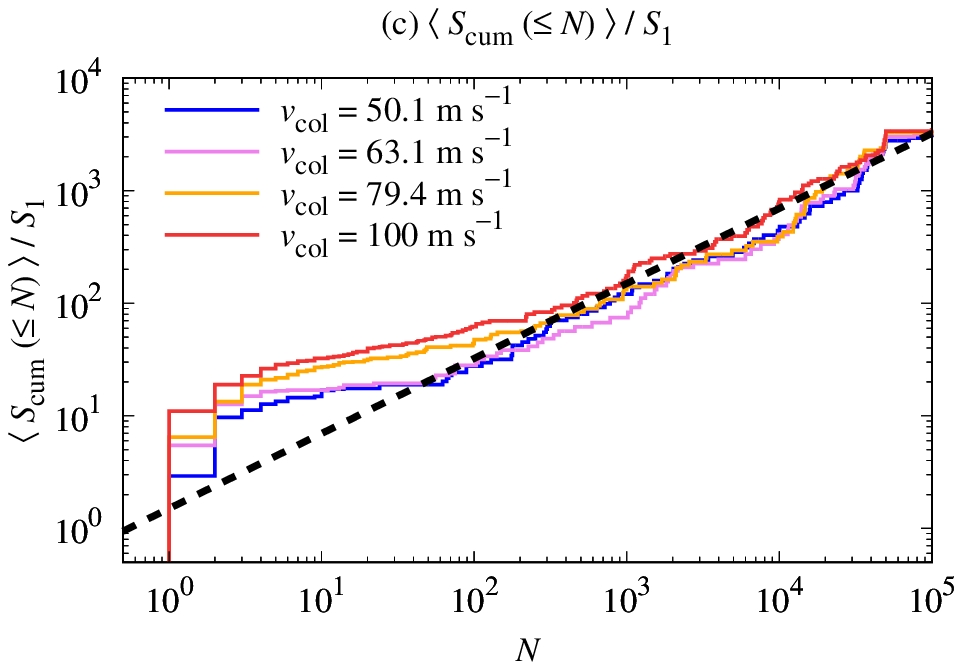}
\caption{
$B_{\rm off}$-weighted average of size distributions of fragments.
(a) ${\langle Z_{\rm cum} (\ge N) \rangle}$.
(b) ${\langle N_{\rm cum} (\le N) \rangle}$.
(c) ${\langle S_{\rm cum} (\le N) \rangle}$.
Black dashed lines are the empirical fittings.
These size distributions are insensitive to $v_{\rm col}$ around $v_{\rm fra}$.
}
\label{fig.size}
\end{figure}

Figures \ref{fig.Zcum}--\ref{fig.Scum} revealed that the size distribution of fragments strongly depends on the strength of viscous dissipation.
When $T_{\rm vis} = 6~{\rm ps}$ is assumed, we found that the size distribution of fragments at collision velocity of $v_{\rm col} \sim v_{\rm fra}$ is approximately fitted by simple curves as shown in Figure \ref{fig.size}.

Here we assume that the number of fragments that contain not less than $N$ and less than $N + {\Delta N}$ particles is given by ${f (N)} {\Delta N}$.
Under this assumption, ${\langle N_{\rm cum} (N) \rangle}$ is given by \begin{equation}
\label{eq.Ndef}
{\langle N_{\rm cum} (\le N) \rangle} \simeq \int_{0}^{N} {\rm d}N~{f (N)} N.
\end{equation}
For arbitrary $N$, the following relation,
\begin{equation}
\label{eq.Nfit}
{\langle N_{\rm cum} (\le N) \rangle} \simeq N,
\end{equation}
reasonably reproduces the numerical results as shown in Figure \ref{fig.size}(b).
Combining Eqs.~{(\ref{eq.Ndef})} and {(\ref{eq.Nfit})}, we obtain the following equation:
\begin{equation}
\label{eq.f}
{f (N)} \simeq \frac{1}{N}.
\end{equation}
Using Eq.~{(\ref{eq.f})}, ${\langle Z_{\rm cum} (\ge N) \rangle}$ and ${\langle S_{\rm cum} (\le N) \rangle}$ are approximately given by
\begin{eqnarray}
{\langle Z_{\rm cum} (\ge N) \rangle} & \simeq & \int_{N}^{N_{\rm tot}} {\rm d}N~{f (N)} \nonumber \\
                                      & \simeq & \log{N_{\rm tot}} - \log{N},
\end{eqnarray}
 and
\begin{eqnarray}
{\langle S_{\rm cum} (\le N) \rangle} / S_{1} & \simeq & \int_{0}^{N} {\rm d}N~{f (N)} N^{2/3} \nonumber \\
                                              & \simeq & \frac{3}{2} N^{2/3},
\end{eqnarray}
respectively.
Black dashed lines in Figures \ref{fig.Zcum}--\ref{fig.size} show the empirical fitting equations described above.
These fittings are in agreement with numerical results.
We note that the size distributions are insensitive to $v_{\rm col}$ around $v_{\rm fra}$.

Here we briefly describe the generalized relations among ${\langle Z_{\rm cum} (\ge N) \rangle}$, ${\langle N_{\rm cum} (N) \rangle}$, and ${\langle S_{\rm cum} (\le N) \rangle}$ as reference.
For simplicity, we assume that ${f (N)}$ is given by a following power-law equation:
\begin{equation}
{f (N)} \simeq c N^{- \alpha} \quad {\left( 1 \le N \le N_{\rm tot} \right)},
\end{equation}
where $c$ and $\alpha$ are constants.
Then ${\langle Z_{\rm cum} (\ge N) \rangle}$, ${\langle N_{\rm cum} (N) \rangle}$, and ${\langle S_{\rm cum} (\le N) \rangle}$ are given as follows:
\begin{align}
{\langle Z_{\rm cum} (\ge N) \rangle} & \simeq   \int_{N}^{N_{\rm tot}} {\rm d}N~{f (N)}                & \nonumber \\
                                      & \simeq   \begin{cases}
                                                    {c {N_{\rm tot}}^{1 - \alpha}} / {( 1 - \alpha )}   & \quad {\left( \alpha < 1 \right)}, \\
                                                     c \log{\left(  N_{\rm tot} / N \right)}            & \quad {\left( \alpha = 1 \right)}, \\
                                                    {c N^{1 - \alpha}} / {( \alpha - 1 )}               & \quad {\left( \alpha > 1 \right)},
                                                 \end{cases}
\end{align}
\begin{align}
{\langle N_{\rm cum} (\le N) \rangle} & \simeq   \int_{1}^{N} {\rm d}N~{f (N)} N            & \nonumber \\
                                      & \simeq   \begin{cases}
                                                    {c N^{2 - \alpha}} / {( 2 - \alpha )}   & \quad {\left( \alpha < 2 \right)}, \\
                                                    c \log{N}                               & \quad {\left( \alpha = 2 \right)}, \\
                                                    {c} / {( \alpha - 2 )}                  & \quad {\left( \alpha > 2 \right)},
                                                 \end{cases}
\end{align}
and
\begin{align}
\frac{\langle S_{\rm cum} (\le N) \rangle}{S_{1}} & \simeq   \int_{1}^{N} {\rm d}N~{f (N)} N^{2/3}         & \nonumber \\
                                                  & \simeq   \begin{cases}
                                                                {c N^{5/3 - \alpha}} / {( 5/3 - \alpha )}  & \quad {\left( \alpha < 5/3 \right)}, \\
                                                                 c \log{N}                                 & \quad {\left( \alpha = 5/3 \right)}, \\
                                                                {c} / {( \alpha - 5/3 )}                   & \quad {\left( \alpha > 5/3 \right)},
                                                             \end{cases}
\end{align}
respectively.
Therefore, largest fragments dominate the mass when $\alpha < 2$, and they also dominate the surface area when $\alpha < 5/3$ is satisfied.
As our numerical results indicate that $\alpha \simeq 1$ for the case of non-zero $T_{\rm vis}$, largest fragments dominate both mass and surface area.

We also note that size distributions of fragments produced by collisions of monolithic rocks (and ice) were studied in laboratory impact experiments.
\citet{1990Icar...87..307M} reviewed these experimental results, and they reported that the size distribution for smaller fragments is approximately given by a power-law with $\alpha \simeq 5/3$, although $\alpha$ weakly depends on the impact velocity and the material parameters.
In addition, when we fit a power-law to the size distribution of larger fragments, $\alpha$ is larger than $5/3$ for catastrophic fragmentation and it strongly depends on the impact velocity and the material parameters.
As $\alpha$ would be associated with the microphysics of fracture propagation, we should investigate what causes the relatively small $\alpha$ of dust aggregates compared with the case of monolithic rocks/ices in future studies.

\section{Discussion}

In Section \ref{sec.vfrag}, we showed that $v_{\rm fra}$ is approximately independent of $T_{\rm vis}$, although $v_{\rm stick}$ strongly depends on $T_{\rm vis}$.
In the following part, we perform the analyses of interparticle velocity and energy dissipation to understand the reason why $v_{\rm fra}$ is insensitive to $T_{\rm vis}$.

\subsection{Normal component of the interparticle velocity}
\label{sec.vrel}

\begin{figure*}
\centering
\includegraphics[width=0.48\textwidth]{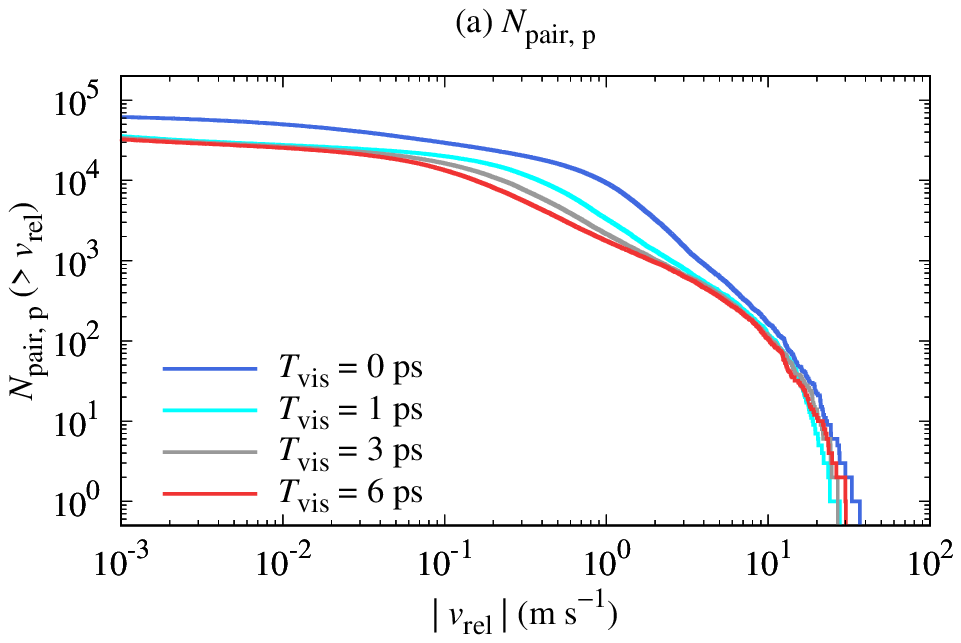}
\includegraphics[width=0.48\textwidth]{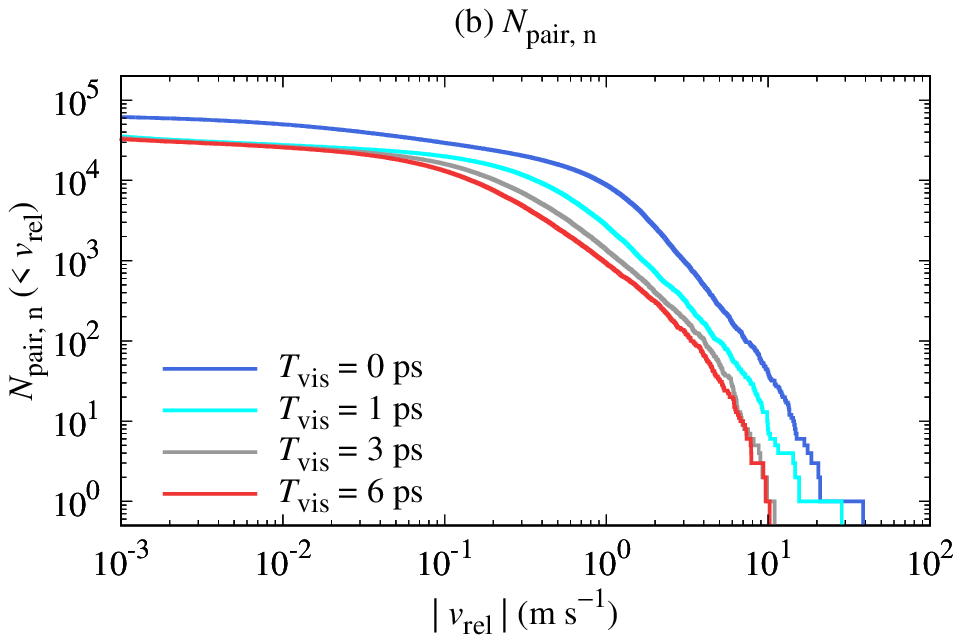}
\caption{
Cumulative frequencies of the normal component of the interparticle velocity, $N_{\rm pair, p} {(> v_{\rm rel} )}$ and $N_{\rm pair, n} {(< v_{\rm rel} )}$, at $t = 0.32~\si{\micro}{\rm s}$ \editX{(see Figures \ref{fig.1}(b), (g), (l), and (q))}.
Here we show the results with $v_{\rm col} = 50.1~{\rm m}~{\rm s}^{-1}$ and ${B_{\rm off}}^{2} = 3/12$.
}
\label{fig.vrel5}
\end{figure*}

\begin{figure*}
\centering
\includegraphics[width=0.48\textwidth]{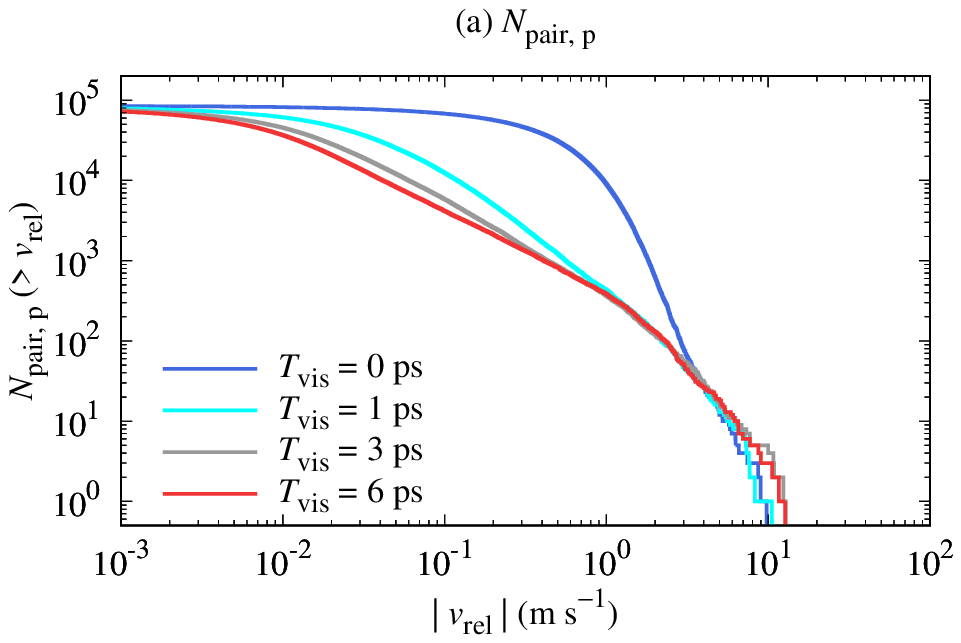}
\includegraphics[width=0.48\textwidth]{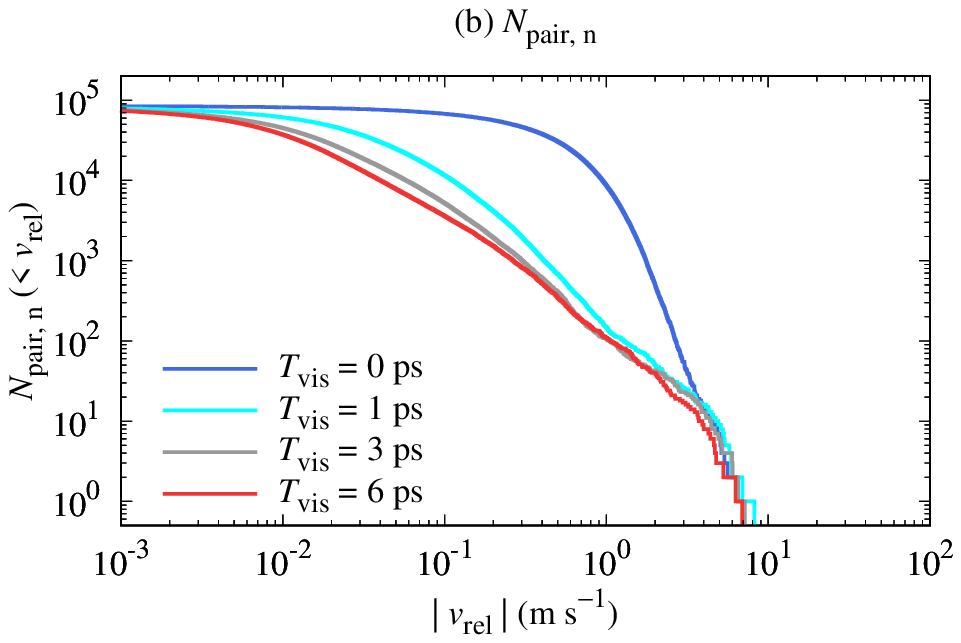}
\caption{
Cumulative frequencies of the normal component of the interparticle velocity, $N_{\rm pair, p} {(> v_{\rm rel} )}$ and $N_{\rm pair, n} {(< v_{\rm rel} )}$, at $t = 0.63~\si{\micro}{\rm s}$ \editX{(see Figures \ref{fig.1}(c), (h), (m), and (r))}.
}
\label{fig.vrel10}
\end{figure*}

Figure \ref{fig.vrel5} shows the normal component of the interparticle velocity, $v_{\rm rel}$, at $t = 0.32~\si{\micro}{\rm s}$ \editX{(see Figures \ref{fig.1}(b), (g), (l), and (q))}.
Here we show the results with $v_{\rm col} = 50.1~{\rm m}~{\rm s}^{-1}$ and ${B_{\rm off}}^{2} = 3/12$.
We calculated $v_{\rm rel}$ for particle pairs whose interparticle distance is less than the threshold distance, $d_{\rm th} = 2.2 r_{1}$.
We set the sign of $v_{\rm rel}$ is positive when two particles approach.
Figure \ref{fig.vrel10} also shows $v_{\rm rel}$ at $t = 0.63~\si{\micro}{\rm s}$ \editX{(see Figures \ref{fig.1}(c), (h), (m), and (r))}.
We define $N_{\rm pair, p} {(> v_{\rm rel} )}$ as the number of pairs whose interparticle velocity is higher than $v_{\rm rel} > 0$, and $N_{\rm pair, n} {(< v_{\rm rel} )}$ as the number of pairs whose interparticle velocity is lower than $v_{\rm rel} < 0$.

We found that both $N_{\rm pair, p} {(> 10~{\rm m}~{\rm s}^{-1} )}$ and $N_{\rm pair, n} {(< - 10~{\rm m}~{\rm s}^{-1} )}$ are less than $10^{3}$ for all cases and are negligibly smaller than $N_{\rm tot}$.
This indicates that viscous energy dissipation, which is proportional to the square of $v_{\rm rel}$, is not the main mechanism for the energy dissipation for collisions between dust aggregates.

\begin{figure}
\centering
\includegraphics[width=\columnwidth]{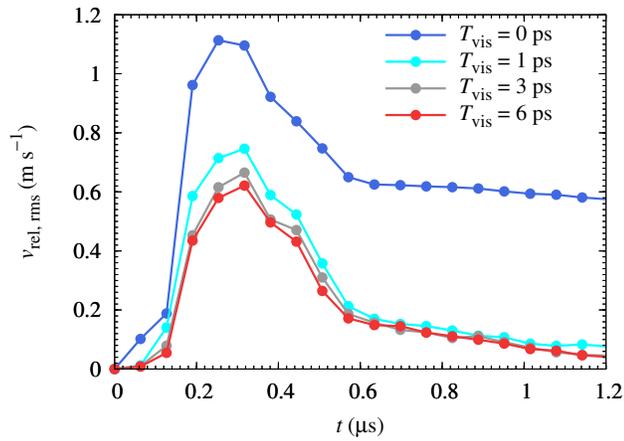}
\caption{
Temporal evolution of the root mean square of $v_{\rm rel}$, $v_{\rm rel, rms}$.
Here we show the results with $v_{\rm col} = 50.1~{\rm m}~{\rm s}^{-1}$ and ${B_{\rm off}}^{2} = 3/12$.
}
\label{fig.vrms}
\end{figure}

The energy dissipation rate at a particle contact due to viscous dissipation force, $\dot{E}_{\rm vis}$, is given by
\begin{eqnarray}
\dot{E}_{\rm vis} & = & - F_{\rm D} v_{\rm rel} \nonumber \\
                  & = & 1.5 \times 10^{-9} \nonumber \\
                  &   & \times {\left( \frac{a}{a_{0}} \right)} {\left( \frac{T_{\rm vis}}{1~{\rm ps}} \right)} {\left( \frac{v_{\rm rel}}{1~{\rm m}~{\rm s}^{-1}} \right)}^{2}~{\rm J}~{\rm s}^{-1}.
\end{eqnarray}
The kinetic energy per one particle with the velocity of $v_{\rm col} / 2$ is
\begin{eqnarray}
\label{eq.E1}
E_{\rm kin, 1} & = & \frac{1}{2} m_{1} {\left( \frac{v_{\rm col}}{2} \right)}^{2} \nonumber \\
               & = & 1.3 \times 10^{-15} {\left( \frac{v_{\rm col}}{50~{\rm m}~{\rm s}^{-1}} \right)}^{2}~{\rm J}.
\end{eqnarray}
Figure \ref{fig.vrms} shows the temporal evolution of the root mean square of $v_{\rm rel}$, $v_{\rm rel, rms}$.
For the cases with non-zero $T_{\rm vis}$, we found that $v_{\rm rel, rms}$ takes the maximum around $t \simeq 0.3~\si{\micro}{\rm s}$ and the maximum value is lower than $1~{\rm m}~{\rm s}^{-1}$.
In addition, $v_{\rm rel, rms}$ decreases with the timescale of $\simeq 0.2~\si{\micro}{\rm s}$ is these cases.
As the total energy dissipation at a particle contact due to viscous dissipation force, $E_{\rm vis}$, is given by the time integration of $\dot{E}_{\rm vis}$, our results indicate that $E_{\rm vis} < E_{\rm kin, 1}$.
In other words, the viscous energy dissipation would not be the main mechanism for the energy dissipation for collisions between dust aggregates.

\subsection{Energy dissipation mechanisms}

\begin{figure*}
\centering
\includegraphics[width=0.48\textwidth]{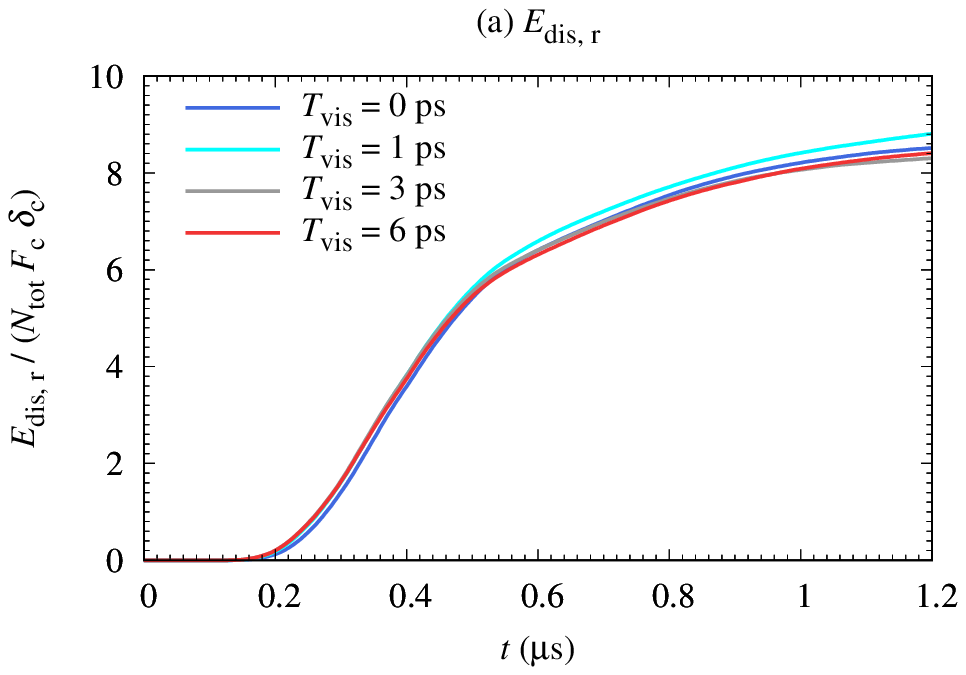}
\includegraphics[width=0.48\textwidth]{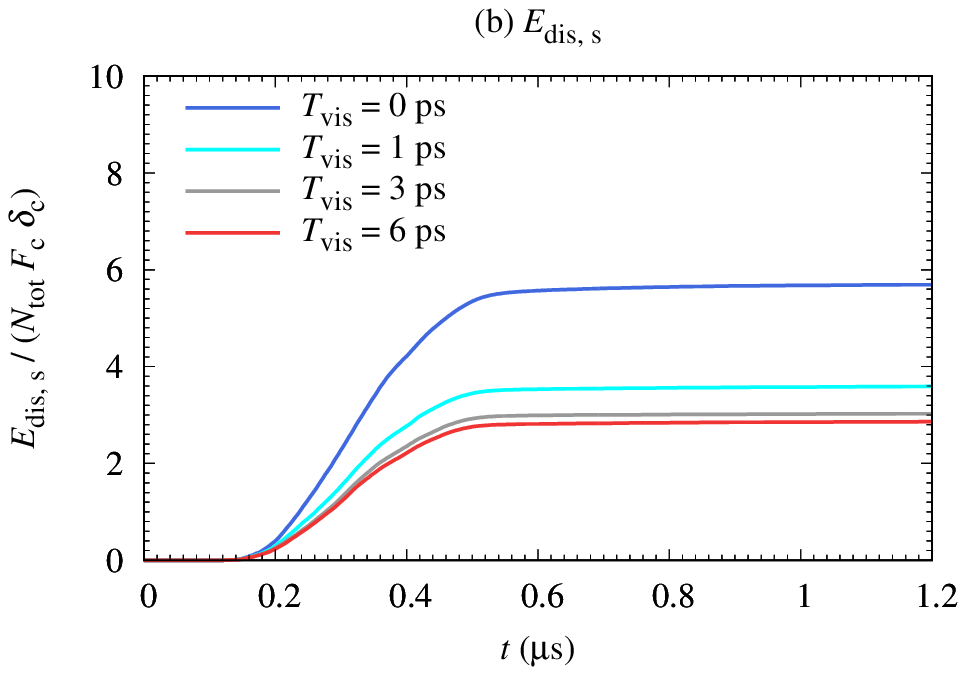}
\includegraphics[width=0.48\textwidth]{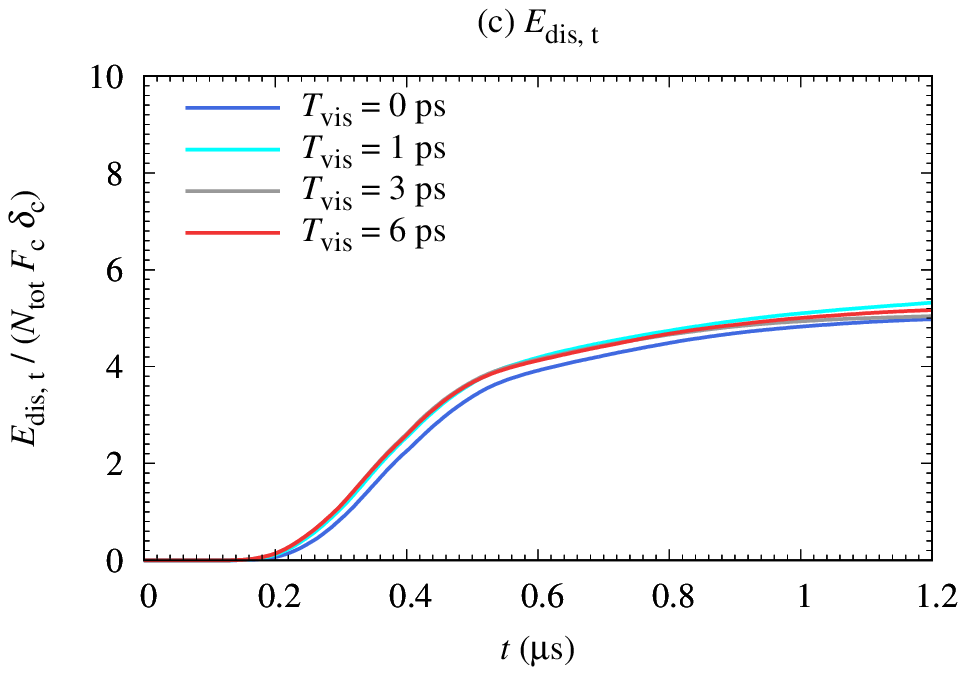}
\includegraphics[width=0.48\textwidth]{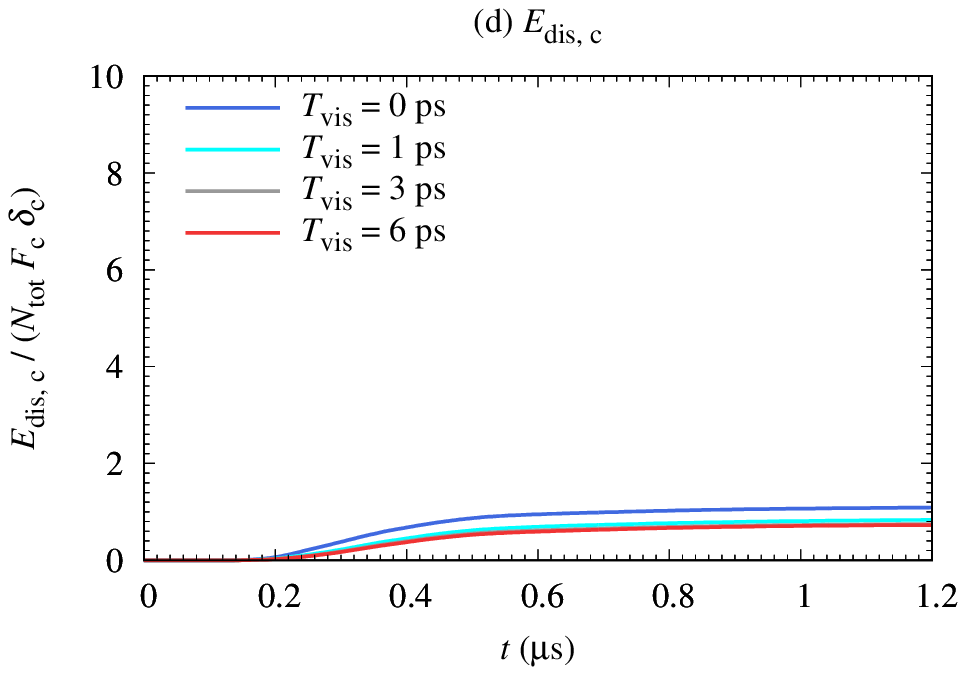}
\includegraphics[width=0.48\textwidth]{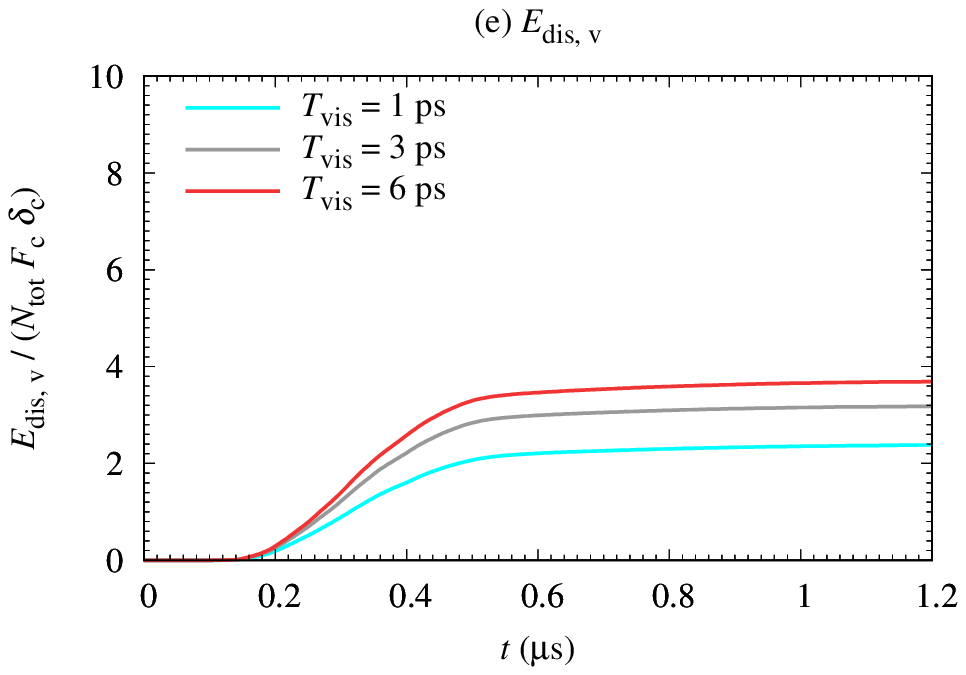}
\caption{
Temporal evolution of the total energy dissipation due to rolling friction ($E_{\rm dis, r}$), sliding friction ($E_{\rm dis, s}$), twisting friction ($E_{\rm dis, t}$), connection and disconnection of particles ($E_{\rm dis, c}$), and viscous drag force ($E_{\rm dis, v}$).
}
\label{fig.edis17}
\end{figure*}

We check the total energy dissipation due to particle interactions from $t = 0$ and discuss which is the main mechanism for the energy dissipation.
Figure \ref{fig.edis17} shows the temporal evolution of the total energy dissipation due to rolling friction ($E_{\rm dis, r}$), sliding friction ($E_{\rm dis, s}$), twisting friction ($E_{\rm dis, t}$), connection and disconnection of particles ($E_{\rm dis, c}$), and viscous drag force ($E_{\rm dis, v}$).
The particle interaction energies for interparticle motions are summarized in Table \ref{table2}.
Here we show the results with $v_{\rm col} = 50.1~{\rm m}~{\rm s}^{-1}$ and ${B_{\rm off}}^{2} = 3/12$.

We found that the main mechanism for the energy dissipation is the rolling friction for all cases shown in Figure \ref{fig.edis17}, and the energy dissipation due to connection and disconnection of particles is the smallest among five interaction mechanisms \editR{(except for the case of $T_{\rm vis} = 0~{\rm ps}$, which results in $E_{\rm dis, v} = 0$)}.
The strength of the energy dissipation due to the viscous drag force increases with increasing $T_{\rm vis}$ \editR{(Figure \ref{fig.edis17}(e))}.
On the other hand, the strength of the energy dissipation due to the sliding friction decreases with increasing $T_{\rm vis}$ \editR{(Figure \ref{fig.edis17}(b))}.
The strength of the energy dissipation due to the twisting friction hardly depends on $T_{\rm vis}$ in this setting \editR{(Figure \ref{fig.edis17}(c))}.
We will further investigate the physical origin of these trends in future studies.

\begin{figure*}
\centering
\includegraphics[width=0.48\textwidth]{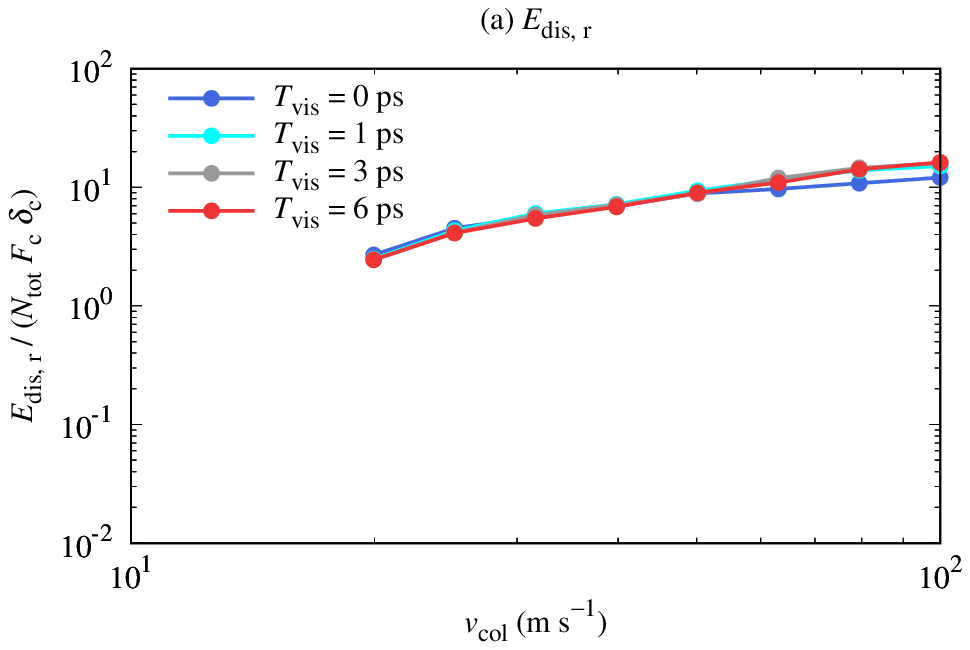}
\includegraphics[width=0.48\textwidth]{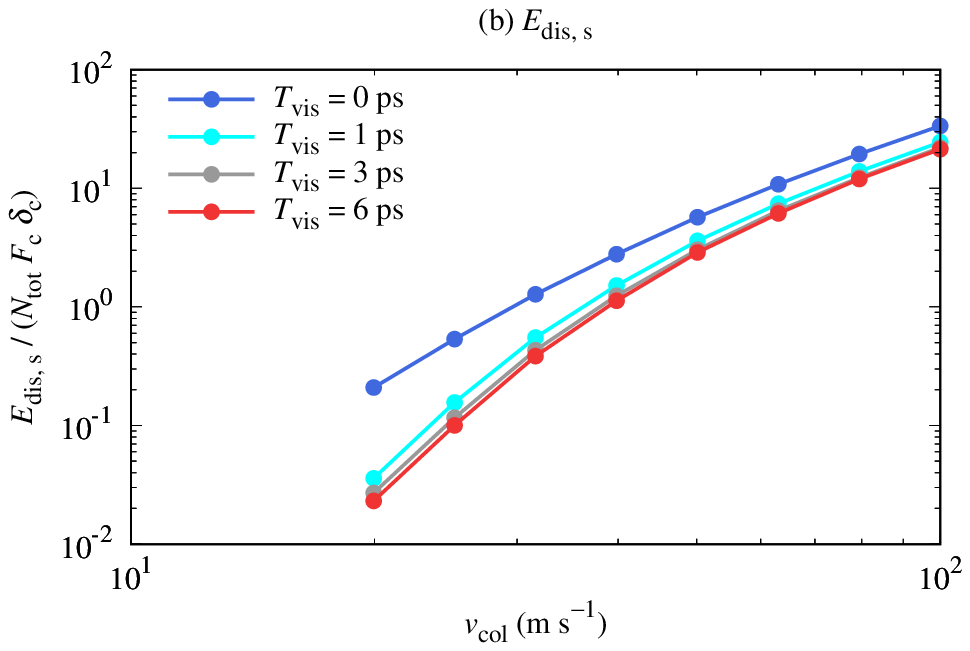}
\includegraphics[width=0.48\textwidth]{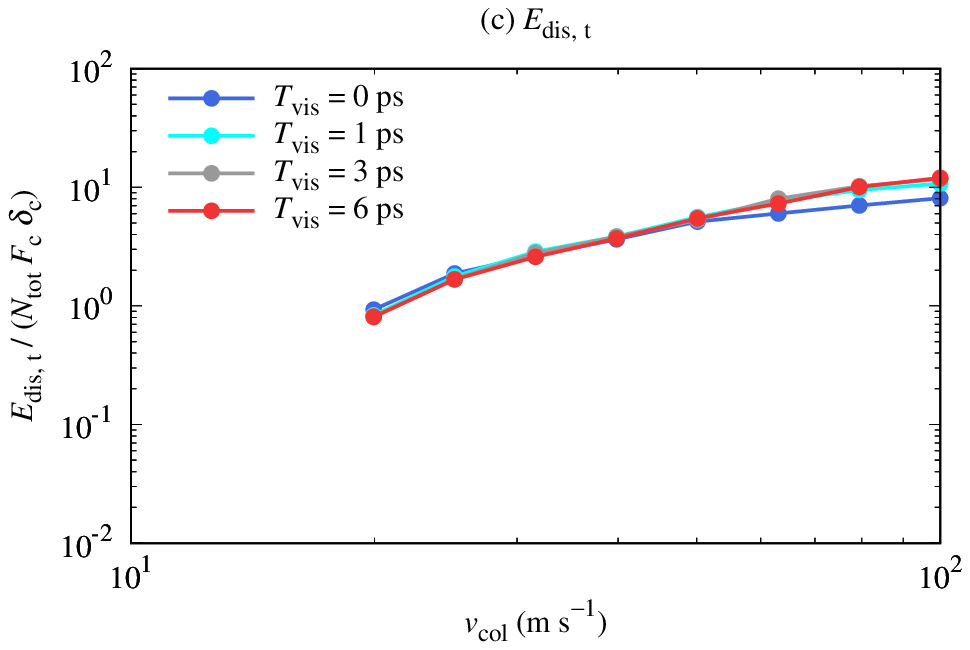}
\includegraphics[width=0.48\textwidth]{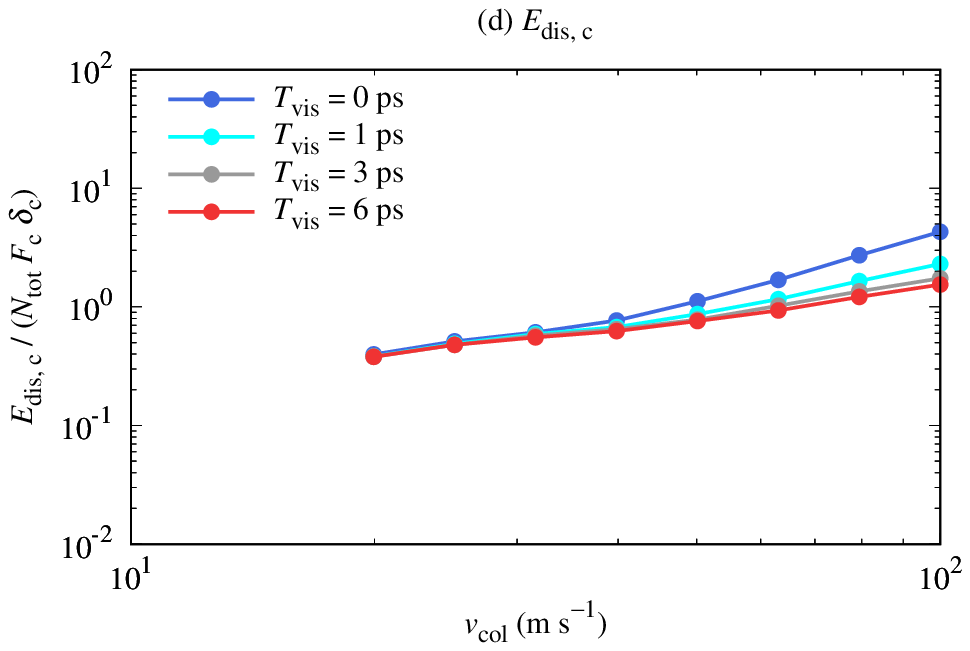}
\includegraphics[width=0.48\textwidth]{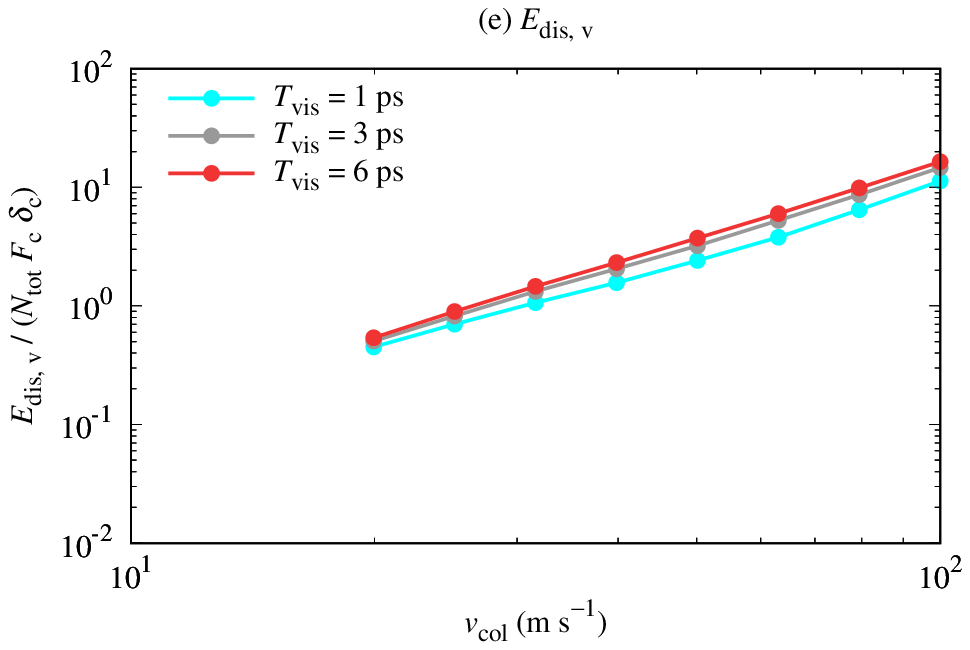}
\caption{
Total energy dissipation due to particle interactions at the end of simulations: (a) $E_{\rm dis, r}$, (b) $E_{\rm dis, s}$, (c) $E_{\rm dis, t}$, (d) $E_{\rm dis, c}$, and (e) $E_{\rm dis, v}$.
\editX{We also show the sum of $E_{\rm dis, s}$ and $E_{\rm dis, v}$ as reference (panel (f)).}
}
\label{fig.edis}
\end{figure*}

Figure \ref{fig.edis} shows the total energy dissipation due to particle interactions at the end of simulations ($E_{\rm dis, r}$, $E_{\rm dis, s}$, $E_{\rm dis, t}$, $E_{\rm dis, c}$, and $E_{\rm dis, v}$).
We set ${B_{\rm off}}^{2} = 3/12$ and investigated the dependence of the total energy dissipation on $T_{\rm vis}$ and $v_{\rm col}$.
We found that \editR{$E_{\rm dis, s}$} strongly depends on $v_{\rm col}$ for all settings of $T_{\rm vis}$ (see Figure \ref{fig.edis}(b)), and \editR{$E_{\rm dis, s}$} is not the main mechanism for the energy dissipation when $v_{\rm col} \lesssim v_{\rm fra}$.
This strong dependence of \editR{$E_{\rm dis, s}$} on $v_{\rm col}$ might be originated from the magnitude relation between $E_{\rm slide}$ and $E_{\rm kin, 1}$ (see Table \ref{table2}).
When $v_{\rm col} \lesssim v_{\rm fra}$, the magnitude relation is $E_{\rm slide} \gg E_{\rm kin, 1}$ and the kinetic energy of colliding aggregates would not be enough to cause large sliding displacement.
In contrast, when $v_{\rm col} \gtrsim 100~{\rm m}~{\rm s}^{-1}$, the condition $E_{\rm slide} \gtrsim E_{\rm kin, 1}$ is achieved and large sliding displacement could take place.
We will test this hypothesis by systematically changing $k_{\rm s}$ in future studies.


The main mechanism for the energy dissipation is $E_{\rm dis, r}$ when the collision velocity is $v_{\rm col} \lesssim v_{\rm fra}$.
We also found that $E_{\rm dis, c} / {( N_{\rm tot} F_{\rm c} \delta_{\rm c} )}$ is on the order of 1 when $v_{\rm col} \sim v_{\rm fra}$ and this is common feature for all cases of $T_{\rm vis}$ (see Figure \ref{fig.edis}(d)).
This is consistent with the fact that the sum of the number of connection and disconnection events is approximately equal to the total number of particles; $N_{\rm con} + N_{\rm cut} \simeq N_{\rm tot}$ (see Section \ref{sec.con}).

We note that \citet{1997ApJ...480..647D} also described the trend for the magnitude relationship of energy dissipation mechanisms.
Although they performed head-on collisions of equal-mass aggregates in two-dimensional space, the trend is similar to our results; the main mechanism for the energy dissipation is rolling for low-speed collisions, and sliding becomes important when $v_{\rm col}$ is large.
We confirmed this trend for head-on collisions in our three-dimensional simulations.

\subsection{Particle connection and disconnection}
\label{sec.con}

\begin{figure*}
\centering
\includegraphics[width=0.48\textwidth]{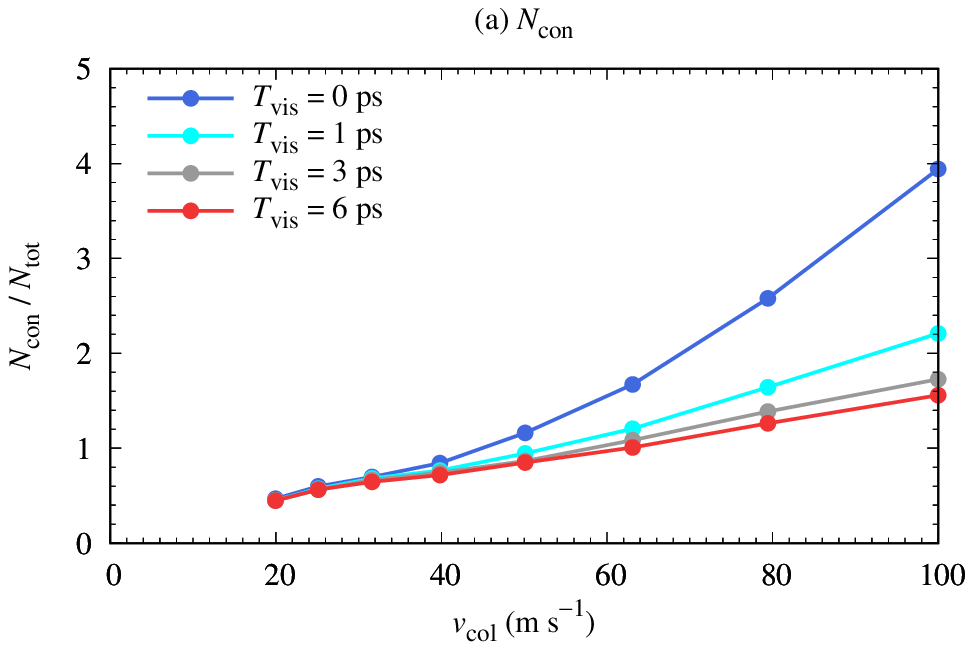}
\includegraphics[width=0.48\textwidth]{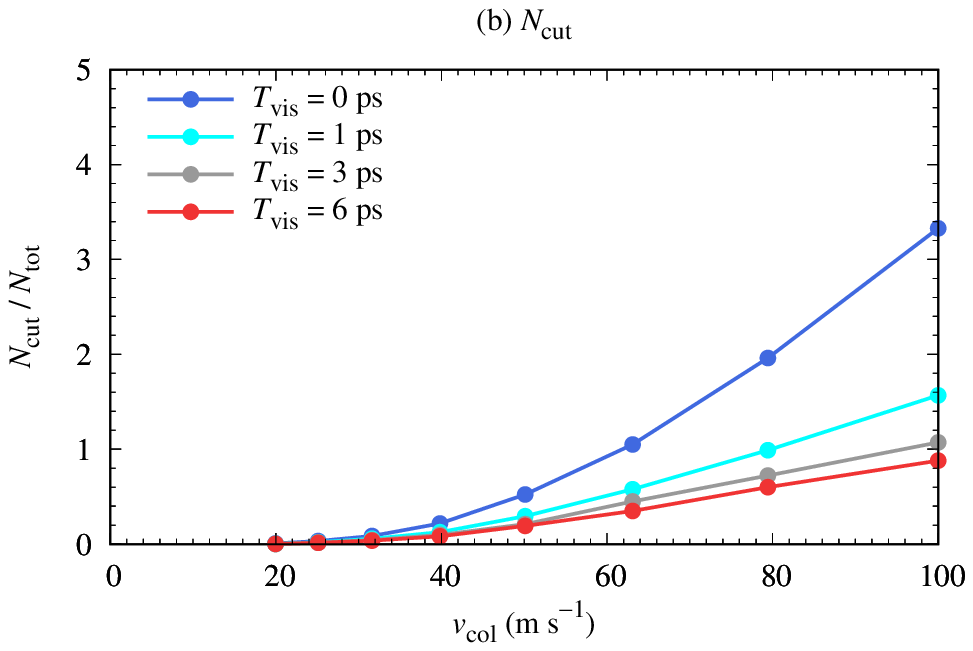}
\caption{
Numbers of connection and disconnection events in a collision between dust aggregates.
(a) Number of connection events, $N_{\rm con}$.
(b) Number of disconnection events, $N_{\rm cut}$.
Here we set ${B_{\rm off}}^{2} = 3/12$.
}
\label{fig.con}
\end{figure*}

Figure \ref{fig.con} shows the number of connection events, $N_{\rm con}$, and disconnection events, $N_{\rm cut}$, in a collision between dust aggregates with ${B_{\rm off}}^{2} = 3/12$.
We found that both $N_{\rm con}$ and $N_{\rm cut}$ increases with decreasing $T_{\rm vis}$ and increasing $v_{\rm col}$.
When we focus on the collision behavior around $v_{\rm col} \simeq v_{\rm fra}$, our numerical simulations revealed that $N_{\rm con} \simeq N_{\rm tot}$ and $N_{\rm cut} < N_{\rm tot}$ for all cases.
This is in agreement with the results shown in Figure \ref{fig.edis}.

\citet{2009ApJ...702.1490W} reported that each particle experiences multiple ($\gg 1$) events of connection and disconnection on average in a head-on collision of dust aggregates (see their Figure 5).
Our results, in contrast, revealed that the average number of events per a particle is approximately one (or a few) in a grazing collision, even if the normalized impact parameter is small (${B_{\rm off}}^{2} = 3/12$).
This would be related to the apparent discrepancy between our results and the results of \citet{umstatter2021scaling}, who reported that the threshold velocity for collisional growth/fragmentation increases with the strength of viscous dissipation force for head-on collisions of dust aggregates.

\section{Summary}

Understanding the collisional behavior of dust aggregates consisting of submicron-sized grains is essential to unveiling how planetesimals form in protoplanetary disks.
It is known that the collisional behavior of individual dust particles strongly depends on the strength of viscous dissipation force \citep[e.g.,][]{2015ApJ...798...34G, 2021ApJ...910..130A}; however, impacts of viscous dissipation on the collisional behavior of dust aggregates have not been studied in detail.

\citet{umstatter2021scaling} reported that the threshold velocity increases with the strength of viscous dissipation force when we consider a head-on collision of dust aggregates.
However, in reality, almost all collision events in protoplanetary disks are oblique collisions, and whether the viscous dissipation helps the collisional growth of dust aggregates is unclear.

In this study, we examined the impacts of viscous dissipation on the collisional behavior of dust aggregates.
We performed \editX{three-dimensional} numerical simulations of collisions between two equal-mass dust aggregates with various collision velocity and impact parameters.
We also changed the strength of viscous dissipation force systematically, which is controlled by $T_{\rm vis}$ called the viscoelastic timescale (see Eq.~{(\ref{eq.Fd})}).
In this study, we set that dust particles constituting dust aggregates are made of water ice, and the all particles have the same radius of $r_{1} = 0.1~\si{\micro}{\rm m}$.
The total number of particles in a simulation is $N_{\rm tot} = 10^{5}$, and both target and projectile aggregates are prepared by ballistic particle--cluster aggregation.
Our key findings are summarized as follows.

\begin{enumerate}
\item{
The threshold velocity for sticking in a head-on collision of individual dust particle ($v_{\rm stick}$) for $T_{\rm vis} = 6~{\rm ps}$ is approximately 10 times higher than that for $T_{\rm vis} = 0~{\rm ps}$.
In contrast, we found that $v_{\rm fra}$ barely depends on $T_{\rm vis}$ within the range of $0~{\rm ps} \le T_{\rm vis} \le 6~{\rm ps}$ (Figure \ref{fig.Vst}).
This indicates that the dominant energy dissipation process for collisions of dust aggregates is different from that for head-on collisions of individual particles.
}
\item{
We analyzed the normal component of the interparticle velocity ($v_{\rm rel}$) from snapshots of our numerical simulations (Figures \ref{fig.vrel5}--\ref{fig.vrms}).
We found that the root mean square of $v_{\rm rel}$ is always lower than $1~{\rm m}~{\rm s}^{-1}$ for non-zero $T_{\rm vis}$, and its decrease timescale is $\simeq 0.2~\si{\micro}{\rm s}$.
From a simple estimate (see Section \ref{sec.vrel}), we confirmed that the viscous energy dissipation would not be the main mechanism for the energy dissipation for collisions between dust aggregates.
}
\item{
We also checked the total energy dissipation due to particle interactions from $t = 0$ and discussed which is the main mechanism for the energy dissipation (Figures \ref{fig.edis17} and \ref{fig.edis}).
We found that the rolling friction is the main mechanism for the energy dissipation when $v_{\rm col} \lesssim v_{\rm fra}$, and the energy dissipation due to connection and disconnection of particles is always not the main mechanism for the energy dissipation.
}
\item{
We showed the size distribution of fragments formed after collisions (Figures \ref{fig.Zcum}--\ref{fig.size}).
We found that largest fragments dominate the mass for all cases in this study.
In addition, with non-zero $T_{\rm vis}$, the largest fragments dominate not only the mass but also the surface area even for $v_{\rm col} = 100~{\rm m}~{\rm s}^{-1}$.
We also obtained a simple empirical fitting for the size distribution of fragments for the case of strong viscous dissipation: ${f (N)} \simeq N^{-1}$ (see Section \ref{sec.emp}).
}
\end{enumerate}

As the energy dissipation due to tangential motions of particles in contact plays a key role in our simulations, we should investigate how collisional outcomes depend on the spring constants and the critical displacements for tangential motions in future studies.
We also plan to perform numerical simulations of collisions between two aggregates made of non-spherical particles to demonstrate how the particle shape affects particle interactions including tangential motions.

\section*{acknowledgments}

\editR{The anonymous reviewer provided a constructive review that improved this paper.}
The authors thank Sebastiaan Krijt and Yukihiko Hasegawa for their helpful comments.
\editR{Numerical computations were carried out on PC cluster at CfCA, NAOJ.}
S.A.\ was supported by JSPS KAKENHI Grant No.\ JP20J00598.
H.T.\ and E.K.\ were supported by JSPS KAKENHI Grant No.\ 18H05438.
This work was supported by the Publications Committee of NAOJ.



\bibliography{sample631}{}

\begin{thebibliography}{}
\expandafter\ifx\csname natexlab\endcsname\relax\def\natexlab#1{#1}\fi
\providecommand{\url}[1]{\href{#1}{#1}}
\providecommand{\dodoi}[1]{doi:~\href{http://doi.org/#1}{\nolinkurl{#1}}}
\providecommand{\doeprint}[1]{\href{http://ascl.net/#1}{\nolinkurl{http://ascl.net/#1}}}
\providecommand{\doarXiv}[1]{\href{https://arxiv.org/abs/#1}{\nolinkurl{https://arxiv.org/abs/#1}}}

\bibitem[{{Adachi} {et~al.}(1976){Adachi}, {Hayashi}, \&
  {Nakazawa}}]{1976PThPh..56.1756A}
{Adachi}, I., {Hayashi}, C., \& {Nakazawa}, K. 1976, Progress of Theoretical
  Physics, 56, 1756, \dodoi{10.1143/PTP.56.1756}

\bibitem[{{Arakawa} \& {Krijt}(2021)}]{2021ApJ...910..130A}
{Arakawa}, S., \& {Krijt}, S. 2021, \apj, 910, 130,
  \dodoi{10.3847/1538-4357/abe61d}

\bibitem[{{Blum} \& {Wurm}(2008)}]{2008ARA&A..46...21B}
{Blum}, J., \& {Wurm}, G. 2008, \araa, 46, 21,
  \dodoi{10.1146/annurev.astro.46.060407.145152}

\bibitem[{{Brisset} {et~al.}(2016){Brisset}, {Hei{\ss}elmann}, {Kothe},
  {Weidling}, \& {Blum}}]{2016A&A...593A...3B}
{Brisset}, J., {Hei{\ss}elmann}, D., {Kothe}, S., {Weidling}, R., \& {Blum}, J.
  2016, \aap, 593, A3, \dodoi{10.1051/0004-6361/201527288}

\bibitem[{{Brisset} {et~al.}(2017){Brisset}, {Hei{\ss}elmann}, {Kothe},
  {Weidling}, \& {Blum}}]{2017A&A...603A..66B}
---. 2017, \aap, 603, A66, \dodoi{10.1051/0004-6361/201630345}

\bibitem[{{Dominik} \& {Tielens}(1995)}]{1995PMagA..72..783D}
{Dominik}, C., \& {Tielens}, A.~G.~G.~M. 1995, Philosophical Magazine, Part A,
  72, 783, \dodoi{10.1080/01418619508243800}

\bibitem[{{Dominik} \& {Tielens}(1996)}]{1996PMagA..73.1279D}
---. 1996, Philosophical Magazine, Part A, 73, 1279,
  \dodoi{10.1080/01418619608245132}

\bibitem[{{Dominik} \& {Tielens}(1997)}]{1997ApJ...480..647D}
---. 1997, \apj, 480, 647, \dodoi{10.1086/303996}

\bibitem[{{Fritscher} \& {Teiser}(2021)}]{2021ApJ...923..134F}
{Fritscher}, M., \& {Teiser}, J. 2021, \apj, 923, 134,
  \dodoi{10.3847/1538-4357/ac2df4}

\bibitem[{{Gundlach} \& {Blum}(2015)}]{2015ApJ...798...34G}
{Gundlach}, B., \& {Blum}, J. 2015, \apj, 798, 34,
  \dodoi{10.1088/0004-637X/798/1/34}

\bibitem[{{G{\"u}ttler} {et~al.}(2010){G{\"u}ttler}, {Blum}, {Zsom}, {Ormel},
  \& {Dullemond}}]{2010A&A...513A..56G}
{G{\"u}ttler}, C., {Blum}, J., {Zsom}, A., {Ormel}, C.~W., \& {Dullemond},
  C.~P. 2010, \aap, 513, A56, \dodoi{10.1051/0004-6361/200912852}

\bibitem[{{G{\"u}ttler} {et~al.}(2012){G{\"u}ttler}, {Hei{\ss}elmann}, {Blum},
  \& {Krijt}}]{2012arXiv1204.0001G}
{G{\"u}ttler}, C., {Hei{\ss}elmann}, D., {Blum}, J., \& {Krijt}, S. 2012, arXiv
  e-prints, arXiv:1204.0001.
\newblock \doarXiv{1204.0001}

\bibitem[{{Hasegawa} {et~al.}(2021){Hasegawa}, {Suzuki}, {Tanaka}, {Kobayashi},
  \& {Wada}}]{2021ApJ...915...22H}
{Hasegawa}, Y., {Suzuki}, T.~K., {Tanaka}, H., {Kobayashi}, H., \& {Wada}, K.
  2021, \apj, 915, 22, \dodoi{10.3847/1538-4357/abf6cf}

\bibitem[{{Hayashi}(1981)}]{1981PThPS..70...35H}
{Hayashi}, C. 1981, Progress of Theoretical Physics Supplement, 70, 35,
  \dodoi{10.1143/PTPS.70.35}

\bibitem[{{Johansen} {et~al.}(2014){Johansen}, {Blum}, {Tanaka}, {Ormel},
  {Bizzarro}, \& {Rickman}}]{2014prpl.conf..547J}
{Johansen}, A., {Blum}, J., {Tanaka}, H., {et~al.} 2014, in Protostars and
  Planets VI, ed. H.~{Beuther}, R.~S. {Klessen}, C.~P. {Dullemond}, \&
  T.~{Henning}, 547, \dodoi{10.2458/azu\_uapress\_9780816531240-ch024}

\bibitem[{{Johnson} {et~al.}(1971){Johnson}, {Kendall}, \&
  {Roberts}}]{1971RSPSA.324..301J}
{Johnson}, K.~L., {Kendall}, K., \& {Roberts}, A.~D. 1971, Proceedings of the
  Royal Society of London Series A, 324, 301, \dodoi{10.1098/rspa.1971.0141}

\bibitem[{{Krijt} {et~al.}(2014){Krijt}, {Dominik}, \&
  {Tielens}}]{2014JPhD...47q5302K}
{Krijt}, S., {Dominik}, C., \& {Tielens}, A.~G.~G.~M. 2014, Journal of Physics
  D Applied Physics, 47, 175302, \dodoi{10.1088/0022-3727/47/17/175302}

\bibitem[{{Krijt} {et~al.}(2013){Krijt}, {G{\"u}ttler}, {Hei{\ss}elmann},
  {Dominik}, \& {Tielens}}]{2013JPhD...46Q5303K}
{Krijt}, S., {G{\"u}ttler}, C., {Hei{\ss}elmann}, D., {Dominik}, C., \&
  {Tielens}, A.~G.~G.~M. 2013, Journal of Physics D Applied Physics, 46,
  435303, \dodoi{10.1088/0022-3727/46/43/435303}

\bibitem[{{Mizutani} {et~al.}(1990){Mizutani}, {Takagi}, \&
  {Kawakami}}]{1990Icar...87..307M}
{Mizutani}, H., {Takagi}, Y., \& {Kawakami}, S.-I. 1990, \icarus, 87, 307,
  \dodoi{10.1016/0019-1035(90)90136-W}

\bibitem[{{Mukai} {et~al.}(1992){Mukai}, {Ishimoto}, {Kozasa}, {Blum}, \&
  {Greenberg}}]{1992A&A...262..315M}
{Mukai}, T., {Ishimoto}, H., {Kozasa}, T., {Blum}, J., \& {Greenberg}, J.~M.
  1992, \aap, 262, 315

\bibitem[{{Musiolik} {et~al.}(2016){Musiolik}, {Teiser}, {Jankowski}, \&
  {Wurm}}]{2016ApJ...827...63M}
{Musiolik}, G., {Teiser}, J., {Jankowski}, T., \& {Wurm}, G. 2016, \apj, 827,
  63, \dodoi{10.3847/0004-637X/827/1/63}

\bibitem[{{Nietiadi} {et~al.}(2022){Nietiadi}, {Rosandi}, {Bringa}, \&
  {Urbassek}}]{2022ApJ...925..173N}
{Nietiadi}, M.~L., {Rosandi}, Y., {Bringa}, E.~M., \& {Urbassek}, H.~M. 2022,
  \apj, 925, 173, \dodoi{10.3847/1538-4357/ac403d}

\bibitem[{{Nietiadi} {et~al.}(2020){Nietiadi}, {Rosandi}, \&
  {Urbassek}}]{2020Icar..35213996N}
{Nietiadi}, M.~L., {Rosandi}, Y., \& {Urbassek}, H.~M. 2020, \icarus, 352,
  113996, \dodoi{10.1016/j.icarus.2020.113996}

\bibitem[{{Nietiadi} {et~al.}(2017){Nietiadi}, {Umst{\"a}tter}, {Tjong},
  {Rosandi}, {Mill{\'a}n}, {Bringa}, \& {Urbassek}}]{2017PCCP...1916555N}
{Nietiadi}, M.~L., {Umst{\"a}tter}, P., {Tjong}, T., {et~al.} 2017, Physical
  Chemistry Chemical Physics (Incorporating Faraday Transactions), 19, 16555,
  \dodoi{10.1039/C7CP02106B}

\bibitem[{{Okuzumi} \& {Hirose}(2012)}]{2012ApJ...753L...8O}
{Okuzumi}, S., \& {Hirose}, S. 2012, \apjl, 753, L8,
  \dodoi{10.1088/2041-8205/753/1/L8}

\bibitem[{{Ormel} \& {Cuzzi}(2007)}]{2007A&A...466..413O}
{Ormel}, C.~W., \& {Cuzzi}, J.~N. 2007, \aap, 466, 413,
  \dodoi{10.1051/0004-6361:20066899}

\bibitem[{{Paszun} \& {Dominik}(2009)}]{2009A&A...507.1023P}
{Paszun}, D., \& {Dominik}, C. 2009, \aap, 507, 1023,
  \dodoi{10.1051/0004-6361/200810682}

\bibitem[{{Poppe} {et~al.}(2000){Poppe}, {Blum}, \&
  {Henning}}]{2000ApJ...533..454P}
{Poppe}, T., {Blum}, J., \& {Henning}, T. 2000, \apj, 533, 454,
  \dodoi{10.1086/308626}

\bibitem[{{Quadery} {et~al.}(2017){Quadery}, {Doan}, {Tucker}, {Dove}, \&
  {Schelling}}]{2017ApJ...844..105Q}
{Quadery}, A.~H., {Doan}, B.~D., {Tucker}, W.~C., {Dove}, A.~R., \&
  {Schelling}, P.~K. 2017, \apj, 844, 105, \dodoi{10.3847/1538-4357/aa7890}

\bibitem[{{Sakurai} {et~al.}(2021){Sakurai}, {Ishihara}, {Furuya}, {Umemura},
  \& {Shiraishi}}]{2021ApJ...911..140S}
{Sakurai}, Y., {Ishihara}, T., {Furuya}, H., {Umemura}, M., \& {Shiraishi}, K.
  2021, \apj, 911, 140, \dodoi{10.3847/1538-4357/abe9ba}

\bibitem[{{Schr{\"a}pler} {et~al.}(2018){Schr{\"a}pler}, {Blum}, {Krijt}, \&
  {Raabe}}]{2018ApJ...853...74S}
{Schr{\"a}pler}, R., {Blum}, J., {Krijt}, S., \& {Raabe}, J.-H. 2018, \apj,
  853, 74, \dodoi{10.3847/1538-4357/aaa0d2}

\bibitem[{{Schr{\"a}pler} {et~al.}(2022){Schr{\"a}pler}, {Landeck}, \&
  {Blum}}]{2022MNRAS.509.5641S}
{Schr{\"a}pler}, R.~R., {Landeck}, W.~A., \& {Blum}, J. 2022, \mnras, 509,
  5641, \dodoi{10.1093/mnras/stab3348}

\bibitem[{{Seizinger} {et~al.}(2013){Seizinger}, {Krijt}, \&
  {Kley}}]{2013A&A...560A..45S}
{Seizinger}, A., {Krijt}, S., \& {Kley}, W. 2013, \aap, 560, A45,
  \dodoi{10.1051/0004-6361/201322773}

\bibitem[{{Shimaki} \& {Arakawa}(2012)}]{2012Icar..221..310S}
{Shimaki}, Y., \& {Arakawa}, M. 2012, \icarus, 221, 310,
  \dodoi{10.1016/j.icarus.2012.08.005}

\bibitem[{{Sirono} \& {Kudo}(2021)}]{2021ApJ...911..114S}
{Sirono}, S.-i., \& {Kudo}, D. 2021, \apj, 911, 114,
  \dodoi{10.3847/1538-4357/abec7c}

\bibitem[{{Sirono} \& {Ueno}(2017)}]{2017ApJ...841...36S}
{Sirono}, S.-i., \& {Ueno}, H. 2017, \apj, 841, 36,
  \dodoi{10.3847/1538-4357/aa6fad}

\bibitem[{{Tanaka} {et~al.}(2012){Tanaka}, {Wada}, {Suyama}, \&
  {Okuzumi}}]{2012PThPS.195..101T}
{Tanaka}, H., {Wada}, K., {Suyama}, T., \& {Okuzumi}, S. 2012, Progress of
  Theoretical Physics Supplement, 195, 101, \dodoi{10.1143/PTPS.195.101}

\bibitem[{{Umst{\"a}tter} \& {Urbassek}(2020)}]{2020A&A...633A..24U}
{Umst{\"a}tter}, P., \& {Urbassek}, H.~M. 2020, \aap, 633, A24,
  \dodoi{10.1051/0004-6361/201936527}

\bibitem[{{Umst{\"a}tter} \&
  {Urbassek}(2021{\natexlab{a}})}]{2021A&A...652A..40U}
---. 2021{\natexlab{a}}, \aap, 652, A40, \dodoi{10.1051/0004-6361/202141581}

\bibitem[{{Umst{\"a}tter} \&
  {Urbassek}(2021{\natexlab{b}})}]{umstatter2021scaling}
---. 2021{\natexlab{b}}, Granular Matter, 23, 1,
  \dodoi{10.1007/s10035-021-01101-w}

\bibitem[{{Wada} {et~al.}(2013){Wada}, {Tanaka}, {Okuzumi}, {Kobayashi},
  {Suyama}, {Kimura}, \& {Yamamoto}}]{2013A&A...559A..62W}
{Wada}, K., {Tanaka}, H., {Okuzumi}, S., {et~al.} 2013, \aap, 559, A62,
  \dodoi{10.1051/0004-6361/201322259}

\bibitem[{{Wada} {et~al.}(2007){Wada}, {Tanaka}, {Suyama}, {Kimura}, \&
  {Yamamoto}}]{2007ApJ...661..320W}
{Wada}, K., {Tanaka}, H., {Suyama}, T., {Kimura}, H., \& {Yamamoto}, T. 2007,
  \apj, 661, 320, \dodoi{10.1086/514332}

\bibitem[{{Wada} {et~al.}(2008){Wada}, {Tanaka}, {Suyama}, {Kimura}, \&
  {Yamamoto}}]{2008ApJ...677.1296W}
---. 2008, \apj, 677, 1296, \dodoi{10.1086/529511}

\bibitem[{{Wada} {et~al.}(2009){Wada}, {Tanaka}, {Suyama}, {Kimura}, \&
  {Yamamoto}}]{2009ApJ...702.1490W}
---. 2009, \apj, 702, 1490, \dodoi{10.1088/0004-637X/702/2/1490}

\bibitem[{{Weidenschilling}(1977)}]{1977MNRAS.180...57W}
{Weidenschilling}, S.~J. 1977, \mnras, 180, 57, \dodoi{10.1093/mnras/180.2.57}

\end{thebibliography}
\bibliographystyle{aasjournal}



\end{document}